%% file: WDMmassfct.publ.tex
\documentclass[useAMS,usenatbib,usedcolumn]{mn2e}
\pdfoutput=1
\usepackage{graphicx}
\usepackage{color}
\input{defs.tex}
\def\lesssim{\la}
\def\gtrsim{\ga}
\def\mwdm{m_{\rm WDM}}

\title[Halo Mass Function and Free Streaming]{Halo Mass Function and the Free Streaming Scale}
\author[Schneider, Smith and Reed]{Aurel Schneider$^{1}$\thanks{Email: aurel.schneider@sussex.ac.uk}, Robert~E.~Smith$^{2}$ and Darren Reed$^{3}$\\
{$^1$Astronomy Centre, Department of Physics and Astronomy, University of Sussex, Brighton, BN1 9QH, UK}\\
{$^2$Max-Planck Institute for Astrophysics, 85740 Garching, Germany}\\
{$^3$Institut de Ci\`encies de l'Espai (IEEC-CSIC), 08193 Bellaterra (Barcelona), Spain}\\
}

\begin{document}


\pagerange{\pageref{firstpage}--\pageref{lastpage}} \pubyear{2012}

\label{firstpage}
\maketitle


\begin{abstract}
  The nature of structure formation around the particle free streaming
  scale is still far from understood. Many attempts to simulate hot,
  warm, and cold dark matter cosmologies with a free streaming cutoff
  have been performed with cosmological particle-based simulations,
  but they all suffer from spurious structure formation at scales
  below their respective free streaming scales -- i.e. where the
  physics of halo formation is most affected by free streaming.  We
  perform a series of high resolution numerical simulations of
  different warm dark matter (WDM) models, and develop an approximate method to subtract
  artificial structures in the measured halo mass function. The
  corrected measurements are then used to construct and calibrate an
  extended Press-Schechter (EPS) model with sharp-$k$ window function
  and adequate mass assignment. The EPS model gives accurate
  predictions for the low redshift halo mass function of CDM and WDM
  models, but it significantly under-predicts the halo abundance at
  high redshifts. By taking into account the ellipticity of the
  initial patches and connecting the characteristic filter scale to
  the smallest ellipsoidal axis, we are able to eliminate this
  inconsistency and obtain an accurate mass function over all
  redshifts and all dark matter particle masses covered by the
  simulations.  As an additional application we use our model to
  predict the microhalo abundance of the standard neutralino-CDM
  scenario and we give the first quantitative prediction of the mass
  function over the full range of scales of CDM structure formation.
\end{abstract}



\maketitle


\section{Introduction}\label{sec:introduction}

The nature of dark matter is one of the major mysteries of modern
physics and a common point of research for particle physics,
astrophysics and cosmology. In the currently favoured cold dark matter
(CDM) model \citep{Peebles1982}, the dark matter particle is supposed
to be a neutralino, the lightest stable particle in supersymmetry
\citep{Jungman1996}. With a mass around 100 GeV, neutralinos decouple
very early and have extremely low thermal velocities, far too low to
influence structure formation on scales relevant for galaxy
formation. As a result we get the common picture of hierarchical
collapse, where large haloes form through mergers of smaller ones, a
process that spans the range from galaxy clusters down to microhaloes
with masses of about the Earth \citep{Hofmann2001,Bertschinger2006}.

One possible alternative to CDM is the warm dark matter (WDM) model,
in which the dark matter particle is a sterile neutrino or gravitino
\citep{BondSzalay1983,Dodelson1994,Colombi1996,Bode2001}. These
particles are much lighter and hence decouple later on, maintaining
their thermal speed and influencing structure formation up to the
scales of dwarf galaxies. While at large scales the collapse in WDM is
hierarchical and identical to CDM, it becomes strongly suppressed
below a characteristic mass scale, where the free streaming of the
particles prevents the haloes to form and the dark matter is
distributed in a smooth background field instead
\citep{SmithMarkovic2011}. Just above this characteristic scale,
haloes form directly through ellipsoidal collapse rather than through
hierarachical growth.

Besides the CDM and WDM models, there are various alternative dark
matter models such as collisional dark matter, where the dark matter
particles have a self interacting force \citep[][and references
  therein]{Vogelsberger2012} or mixed dark matter, which consists of a
mixture of cold and warm particles
\citep{Maccio2012,Anderhalden2012,Anderhalden2012b}. All these models
are indistinguishable from CDM at large scales and produce modified
clustering below some characteristic scale.

The nonlinear structure formation of a CDM universe without free
streaming has been studied extensively and to high accuracy with high
resolution cosmological simulations \citep{Davis1985,Springel2005} as
well as analytical and semi-analytical approaches
\citep{Press1974,Bond1991,ShethTormen1999}. However, as soon as free
streaming effects are involved, both numerical simulation and
analytical models fail to predict the correct halo abundances at low masses. 
Instead of a strong suppression, they produce large
numbers of haloes at small masses. In the case of numerical
simulations, the failure can be attributed to the fact that, if we
represent the density field through a discrete set of particles, then
around each particle there will be a local gravitational sink. This
sink can attract other particles and can trigger the collapse of
`artificial' haloes even in the absence of cosmological perturbations
\citep{Wang2007}. On the other hand, analytical approaches fail
because they have a strong dependence on the adopted smoothing scale
where the linear power drops to zero \citep{Bertschinger2006,
  Schneider2012}. Very recently there have been attempts to cure these
problems with new simulation techniques \citep{Hahn2012} as well as a
modification of the extended Press-Schechter (EPS) approach
\citep{Benson2012}. Whilst both of these approaches are promising,
they have not demonstrated that they are uniquely convergent, nor do
they currently reproduce the correct structure formation for small
mass scales.

In this work, we study the halo mass function in the presence of WDM
free streaming, using both high resolution numerical simulations and
the EPS approach. We look at the effects of the artificial clumping in
our simulations and propose a simple approximative method to subtract
artificial haloes in the mass function. After this correction the
measured mass function exhibits the expected turnover and steep
decrease towards small masses. In a second step we apply an EPS
approach with adequate filtering and mass assignment, which recovers
the downturn of the mass function and gives a good match to the
corrected measurements from our simulations.

As an additional application of our EPS recipe, we predict, for the
first time, the neutralino-CDM mass function over the entire halo mass
range, from the largest clusters down to the smallest earth-mass sized
microhaloes.

The paper is structured as follows: In \S2 we take a general look at
the free streaming and its effect on the linear power spectrum as well
as the role of the late time thermal velocities. \S3 is devoted to the
numerical simulations of different WDM cosmologies and the difficulty
of artificial halo formation. In \S4 we derive a model for the mass
function with appropriate mass assignment and compare it to our
simulations. This includes a method to correct for the ellipticity of
initial patches in a spherically averaged Gaussian field. Finally, we
apply our method to predict the mass function of a neutralino-CDM
cosmology in \S5, and we conclude in \S6.


\section{The Free Streaming Scale}\label{sec:freestreaming}

The thermal velocities of the dark matter particles have a direct
influence on structure formation, since they tend to erase primordial
perturbations below a certain scale. This scale depends on the mass of
the dark matter particle as well as on its formation mechanism.

Usually the effect of the free streaming is quantized by the length a
particle travels before the primordial perturbations start to grow
substantially, which happens to be around matter-radiation
equality. This approximate calculation leads to the
free-streaming length
\be \lambda_{\rm fs} = \int_{0}^{t_{\rm EQ}} \frac{v(t)dt}{a(t)} 
\approx  \int^{t_{\rm NR}}_{0}
\frac{cdt}{a(t)} + \int_{t_{\rm NR}}^{t_{\rm EQ}} \frac{v(t)dt}{a(t)} \ ,
\ee
where $t_{\rm NR}$ is the epoch when the dark matter particles become
non-relativistic, which occurs as soon as ${T_{\rm DM}<m_{\rm
 DM}c^2/3k_{\rm B}}$. Here we have introduced the scale factor 
$a$, the mass of the dark matter particle $m_{\rm DM}$, and its 
characteristic temperature $T_{\rm DM}$. In the relativistic case, 
the mean peculiar velocity of the particle is simply $v(t)\sim c$. In the 
non-relativistic regime its momentum simply redshifts with the expansion: 
$v\propto a(t)^{-1}$. This leads to
\be \lambda_{\rm fs} \approx  r_{\rm H}(t_{\rm NR})
\left[1+\frac{1}{2}\log \frac{t_{\rm EQ}}{t_{\rm NR}}\right]
\label{freestreaminglength}\ ,
\ee
where $r_{\rm H}(t_{\rm NR})$ is the comoving size of the horizon at
$t_{\rm NR}$. 
Below the free streaming length $\lambda_{\rm fs}$ all
perturbations are wiped out, the dark matter particles being in a
smooth background density field instead.

An alternative way of understanding the effects of free streaming can
be obtained by following the critical Jeans mass through cosmic
history. The linear evolution of a total matter perturbation 
may be expressed as
\be\label{eqpert}
\frac{d^2\delta}{dt^2} + 2H(t)\frac{d\delta}{dt} = 
\left[4\pi G \rhob(t) - \frac{\sigma_{\rm v}^2(t)k^2}{a^2}\right]\delta \ ,
\ee
where $\delta({\bf r},t)=[\rho({\bf r},t)-\rhob(t)]/\rhob(t)$ is 
the matter density perturbation, $\rhob(t)$ is the background 
density of the Universe, $\sigma_{\rm v}$ is the dark matter
velocity dispersion, and $H(t)\equiv\dot{a}/a$ is the expansion
rate. This expression holds on scales well below the horizon (and
for non relativistic species). It may be noted that a necessary
condition for growing mode solutions is that the right-hand-side of
this equation stays positive. This leads one to introduce the
effective Jeans mass
\be\label{Jeansmass}
M_{\rm J}(t)=\frac{4\pi}{3}\rho_m(t)\left[\frac{\pi}{k_J}\right]^3
=\frac{4\pi}{3}\rho_{\rm m}(t)\left[\frac{\pi \sigma_{\rm v}^2(t)}{4G\rhob(t)}\right]^{3/2} \ .
\ee
For $M<M_{\rm J}$ perturbations will be damped. Note that
\Eqn{Jeansmass} depends on the dark matter density $\rho_{\rm m}$ as
well as on the background density $\rhob$, which includes all
cosmic components. The two densities evolve differently, since dark
matter becomes non relativistic well before matter radiation
equality.


\begin{figure}
\centering{ 
  \includegraphics[width=8.5cm]{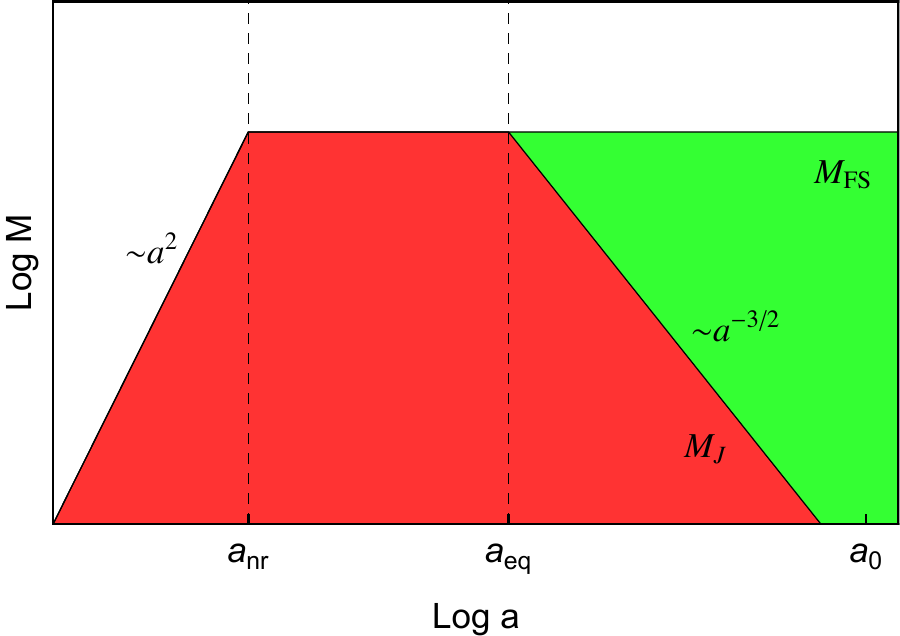}}
\caption{\small{Schematic representation of the Jeans mass M$_J$ and
    the free streaming mass M$_{\rm FS}$ as they evolve through cosmic
    history. In the red coloured region the density perturbations are
    not growing because they are Jeans stable. The green coloured
    region is Jeans unstable but the perturbations are completely
    wiped out due to the velocity free streaming. The free streaming
    scale corresponds to the maximum value of the Jeans
    mass.}}\label{Jeansevolution}
\end{figure}


With the help of \Eqn{Jeansmass}, it is now straight forward to
qualitatively trace the evolution of the Jeans criterion through
cosmic history: In the very early universe the dark matter component
is still relativistic ($\sigma_{\rm v} \sim c$) and the Jeans mass is
growing. As soon as the dark matter becomes non relativistic, its
thermal velocity dispersion cools in the Hubble flow ($\sigma_{\rm v}\sim
a^{-1}$) and the Jeans mass becomes approximately constant; during
this still radiation-dominated era, $\rhob \sim a^{-4}$ while
$\rho_{\rm m} \sim a^{-3}$. This is the case until about
matter-radiation equality, where the evolution of the background
density changes and the Jeans mass drops significantly.

As soon as a mass scale becomes Jeans stable, all perturbations below
this mass scale are damped to zero. The free streaming scale can
therefore alternatively be defined as the maximum value of the Jeans
mass in cosmic history. This is usually the case around
matter-radiation equality, where the Jeans scale happens to be of the
order of the free streaming scale defined in
\Eqn{freestreaminglength}.

The evolution of the Jeans mass and the corresponding free streaming
mass is summarized in \Fig{Jeansevolution}, where the red coloured area
represents scales with strictly no growing solutions, while the green
coloured region stands for scales with growing solutions but erased
initial perturbations. While the Jeans and the free streaming mass are
comparable until matter-radiation equality, they differ substantially
in scale today.


\subsection{Power spectrum}\label{sec:powerspectrum}

A detailed calculation of the effects of the free streaming on the
power spectrum of initial perturbations can only be obtained by
solving the coupled linearized Einstein-Boltzmann equations for all
relevant species in the Universe. The result is usually given in terms
of a transfer function, which is a mapping of the primordial
perturbations from the end of inflation to the moment when the first
perturbations become nonlinear.

In the case of a WDM universe several transfer functions have been
proposed, taking into account different production mechanisms for the
dark matter particle. In the following we use the formula presented in
\citet{Viel2005}:
\be\label{TFwdm}
T_{\rm WDM}(k)=\left[\frac{P_{\rm lin}^{\rm WDM}}{P_{\rm lin}^{\rm CDM}}\right]^{1/2}
=\left[1+(\alpha k)^{2 \mu}\right]^{-5/\mu},
\ee
with $\mu=1.12$ as well as 
\be 
\alpha = 0.049 \left[\frac{m_{\rm WDM}}{\keV}\right]^{-1.11}
\left[\frac{\Omega_{\rm WDM}}{0.25}\right]^{0.11}\left[\frac{h}{0.7}\right]^{1.22}\Mpc,
\ee
which holds for a thermally produced dark matter candidate. A direct
translation to the mass of a sterile neutrino is given by the fitting
function
\be\label{eq:massrescale}
m_{\nu_s}=4.43 \keV \left(\frac{m_{\rm WDM}}{1 \keV}\right)^{4/3}
\left(\frac{\Omega_{\rm WDM}}{0.1225}\right)^{-1/3}.
\ee

The WDM free streaming introduces a characteristic scale of
suppression, and it is convenient to define it to be the `half-mode'
scale at which the WDM transfer function drops to $1/2$
\citep{Schneider2012}. The half-mode mass scale is given by
\be\label{hmmassWDM}
M_{\rm WDM}^{\rm hm}=\frac{4\pi}{3}\rhob
\left[\pi\alpha\left(2^{\mu/5}-1\right)^{-\frac{1}{2\mu}}\right]^3.
\ee
and is about at the mass scale of a dwarf galaxy, depending on the exact
WDM particle mass.

In the case of a CDM cosmology with a neutralino dark matter
candidate, the free streaming cutoff scale is much smaller. The
transfer function obtained by \citet{Green2004} has the form 
\be\label{TFcdm}
T_{\rm N}(k)=\left[1-\frac{2}{3}\left(\frac{k}{k_{\rm A}}\right)^2\right]
\exp\left[-\left(\frac{k}{k_{\rm A}}\right)^2-\left(\frac{k}{k_{\rm B}}\right)^2\right]
\ee
where
\ba\label{kA}
k_{\rm A} & = & 
2.4\times10^6\left(\frac{m_{\rm N}}{100\rm GeV}\right)^{1/2} \nn \\
& & \times  \frac{\left(T_{\rm kd}/30 \rm MeV\right)^{1/2}}{1 + \log(T_{\rm kd}/30 \rm MeV)/19.2} [\kMpc]\ ,
\ea
\be 
k_{\rm B} = 5.4\times10^7\left(\frac{m_{\rm N}}{100\rm GeV}\right)^{1/2}\left(\frac{T_{\rm kd}}{30 \rm MeV}\right)^{1/2} [\kMpc],\label{kB}
\ee
where $T_{\rm kd}$ is the kinetic decoupling temperature, which
depends on the specific parameters of the supersymmetric
model. Typical values for $T_{\rm kd}$ vary between 20 MeV and 35 MeV,
but more extreme values are possible \citep[see][for more
  details]{Green2005}.

Owing to the fact that \Eqn{TFcdm} is not algebraically solvable for
$k$, an expression for the half-mode scale cannot be given. However,
for a 100 GeV neutralino with a decoupling temperature of 30 MeV the
half-mode mass is $M=2.9\times 10^{-6} \Msol$, which roughly
corresponds to the mass of the Earth.

\begin{table*}
   \centering{\label{simtable}
\begin{tabular}{|l|c|c|c|c|c|c|c|}
  \hline
  Sim label & $L\, [\Mpc]$  & $N_{\rm sim}$ &  RS & $m_{\rm WDM}\, [\keV]$ & $M_{\rm hm}[\Msol]$ & $m_p \, [\Msol]$ & $l_{\rm soft} [\kpc]$ \\
  \hline
  CDM\_L256  &  256  & $1024^3$ &  345897  & $\infty$ & 0 &  $1.18 \times 10^{9}$   &  5.00 \\
  CDM\_L64    &    64  & $1024^3$ &  345897  & $\infty$ & 0  & $1.83 \times 10^{7}$   &  1.25 \\
  CDM\_L16    &    16  & $1024^3$ &  345897  & $\infty$ & 0  & $2.86 \times 10^{5}$   &  0.31 \\
  \hline
  WDM\_m1.0\_L256 &  256  & $1024^3$ &  345897  & 1.0  &  $1.3\times10^{10}$  &  $1.18 \times 10^{9}$  &  5.00 \\
  WDM\_m1.0\_L64   &    64  & $1024^3$ &  345897  & 1.0  &  $1.3\times10^{10}$  &  $1.83 \times 10^{7}$  &  1.25 \\
  WDM\_m1.0\_L16a   &    16  & $1024^3$ &  345897  & 1.0  &  $1.3\times10^{10}$  &  $2.86 \times 10^{5}$  &  0.31 \\
  WDM\_m1.0\_L16b   &    16  & $1024^3$ &  234786  & 1.0  &  $1.3\times10^{10}$  &  $2.86 \times 10^{5}$  &  0.31 \\
  WDM\_m1.0\_L16c   &    16  & $1024^3$ &  123675  & 1.0  &  $1.3\times10^{10}$  &  $2.86 \times 10^{5}$  &  0.31 \\
  \hline
  WDM\_m0.5\_L256  & 256  & $1024^3$ &  345897  & 0.5  &  $1.3\times10^{11}$  &  $1.18 \times 10^{9}$  &  5.00 \\
  WDM\_m0.5\_L64    &   64  & $1024^3$ &  345897  & 0.5  &  $1.3\times10^{11}$  &  $1.83 \times 10^{7}$  &  1.25 \\
  \hline
  WDM\_m0.25\_L256 & 256   & $1024^3$ &  345897  & 0.25 &  $1.3 \times 10^{12}$  &  $1.18 \times 10^{9}$  & 5.00 \\
  WDM\_m0.25\_L64a &   64   & $1024^3$ &  345897  & 0.25 &  $1.3 \times 10^{12}$  &  $1.83 \times 10^{7}$  & 1.25 \\
  WDM\_m0.25\_L64b &   64   & $1024^3$ &  234786  & 0.25 &  $1.3 \times 10^{12}$  &  $1.83 \times 10^{7}$  & 1.25 \\
  WDM\_m0.25\_L64c &   64   & $1024^3$ &  123675  & 0.25 &  $1.3 \times 10^{12}$  &  $1.83 \times 10^{7}$  & 1.25 \\
  WDM\_m0.25\_L64d &   64   & $1024^3$ &  012564  & 0.25 &  $1.3 \times 10^{12}$  &  $1.83 \times 10^{7}$  & 1.25 \\
  \hline
\end{tabular}}
\caption{\small{Numerical simulations used in this paper.  Columns
    from left to right: simulation name; simulation box-size ($L$);
    particle number ($N_{\rm sim}$); random seed number (RS); mass of
    WDM particle ($m_{\rm WDM}$); half-mode mass-scale ($M_{\rm hm}$);
    mass of simulation particles $(m_{\rm p})$; comoving softening
    length ($l_{\rm soft}$). The simulations with $L=256 \Mpc$ have
    already been published in
    \citet{Schneider2012}. \label{simtable}}}
\end{table*}


\subsection{The role of late time velocities}

Numerical simulations of cosmologies with a non negligible free
streaming length are usually done by simply assuming a cutoff in the
initial power spectra (as discussed above) without directly including
thermal velocities in the numerical simulations. This approach has
been applied for simulations in WDM
\citep{Bode2001,Lovelletal2011,Schneider2012} as well as simulations
of tiny high redshift boxes in CDM \citep{Diemand2005,
  Ishiyama2010,Anderhalden2013}. Neglecting the late time thermal
velocity contribution is a good approximation as long as the Jeans
mass is well below the mass resolution at the initial redshift of the
simulation.

After matter-radiation equality at $z_{\rm eq}\sim3200$, the
overdensities grow significantly while the Jeans mass drops with a
rate of $M_{\rm J}\sim a^{-3/2}$ (as discussed in
\S\ref{sec:freestreaming} and in the corresponding
Fig.~\ref{Jeansevolution}). This means that already at $z\sim600$ the
Jeans mass is more than one order of magnitude and at $z\sim100$ more
than two orders of magnitude below the free streaming scale. For
simulations with a typical starting redshift of $z\sim100$, it is
therefore impossible to see a direct effect of initial particle
thermal velocities on the WDM mass function -- the artificial clumping
scale would be orders of magnitude larger than the thermal velocity
Jeans scale at the initial epoch of the simulation. This also 
holds for halo density profiles, where the effects of thermal 
velocities (transformation of halo cusps into central cores) are 
only observable if the late time velocities are artificially boosted 
\citep{Villaescusa-Navarro2011,Maccio2012b,Shao2013}.

\citet{Benson2012} included the effect of late time thermal velocities
into their mass function calculation and found a very prominent
effect, which changes the shape of the mass function at scales around
the cutoff. Their model is based on work of \citet{Barkana2001}, who
included the effect of velocities in an isolated simulation of
spherical collapse. The starting redshift of the Barkana et
al. simulation, however, is at matter-radiation equality, at a time
where all relevant perturbations are still extremely small and deep in
the linear regime. Their spherical collapse simulation can therefore
be understood as a simplified method to solve the linear Boltzmann
equation \citep[where velocities have been included as well; see for
  example][]{Viel2005} and, not surprisingly, leads to a cutoff at
about the same scale as the cutoff in the WDM transfer
function. Starting the spherical collapse much later at a redshift
just before the relevant modes become nonlinear, would strongly
reduce the influence of the thermal velocities and would push the
effect on the mass function to much smaller scales, orders of
magnitude below the relevant half-mode scale of \Eqn{hmmassWDM}.


\section{Simulating the WDM universe}\label{sec:simulations}
In this section we present our set of WDM simulations and give details
about the initial conditions, the gravity code, and the
characteristics of the individual runs. In a second part we then
discuss the issue of artificial clumping in detail and develop a way
to deal with it for the purpose of the halo mass function.


\subsection{Characteristics of the simulations}
For all our simulations we used a WMAP7 cosmology with the parameters
$\Omega_{\rm m}=0.2726$, $\Omega_{\Lambda}=0.7274$, $\Omega_{\rm
  b}=0.046$, $h=0.704$, $n_s=0.963$, and $\sigma_8=0.809$
\citep{Komatsuetal2011short}. The CDM transfer function was generated
with the {\tt CAMB} code of \citet{Lewisetal1999}. For the initial
conditions we used the {\tt 2LPT} code
\citep{Scoccimarro1998,Crocceetal2006} with an initial redshift of
$z_{\rm IC}=49$ for runs with box size $L=256\,\Mpc$ and $z_{\rm
  IC}=99$ for runs with box sizes $L=64\,\Mpc$ and $L=16\,\Mpc$. All
simulations have been performed with {\tt PKDGRAV}, a treecode with
high order multipole expansion and adaptive timestepping
\citep{Stadel2001}.

A summary of all simulations including some important physical and
numerical quantities is listed in Table~\ref{simtable}. All
simulations have $1024^3$ particles. The simulations with boxsize
$L=256\,\Mpc$ have already been published in \citet{Schneider2012},
the others have been performed during the last year on the SuperMUC
cluster in Munich and the CSCS cluster in Lugano.  For the instructive
purposes of resolving halo formation well below the cutoff scale, we
focus on cosmologies somewhat warmer than the canonical 2 keV WDM
candidate.

The halo finding was done using a friends-of-friends (FoF) algorithm
\citep{Davis1985} provided by the {\tt $N$-Body Shop}\footnote{\tt
  www-hpcc.astro.washington.edu/tools/fof.html}, with the usual
linking length of $b=0.2$ and with no unbinding of haloes.

For the smallest boxes of $L=16\,\Mpc$ we performed a finite volume
correction to compensate for the missing large-power modes. This was
done in the simplest way by truncating the integrals present in
\Eqns{var}{dvar} at scales larger than the box length, taking the
ratio of the truncated to non-truncated mass function and multiplying
this ratio to the simulation measurements \citep[see for
  example][]{Watson2012}.  Whilst this simple correction neglects the
effects of the discrete Fourier mode distribution and the run to run
sample variance of each realization, the resulting corrected mass
function is similar to that obtained using the correction technique of
\citet{Reed2007}.  The result is an increase in the abundance of
haloes at large mass scales, and, somewhat counter intuitively, a
decrease in the abundance of haloes for smaller mass scales. For the
larger boxes $L=\{64,256\}\,\Mpc$, no finite volume correction has
been implemented, since it has an negligible influence on the halo
abundance.


\subsection{Artificial haloes and resolution}

Numerical simulations of cosmologies with truncated initial power
spectrum produce artificial clumping at scales beyond the cutoff. In
the simulation outputs these artefacts are most visible in cosmic
filaments, where they form equally spaced clumps that are strongly
resolution dependent \citep{Wang2007,Schneider2012}. They seem however
to be present in all environments from underdense voids to high
density clusters and therefore cannot simply be cut out of the
simulation analysis.

\begin{figure}
\centering{ 
  \includegraphics[width=8.5cm]{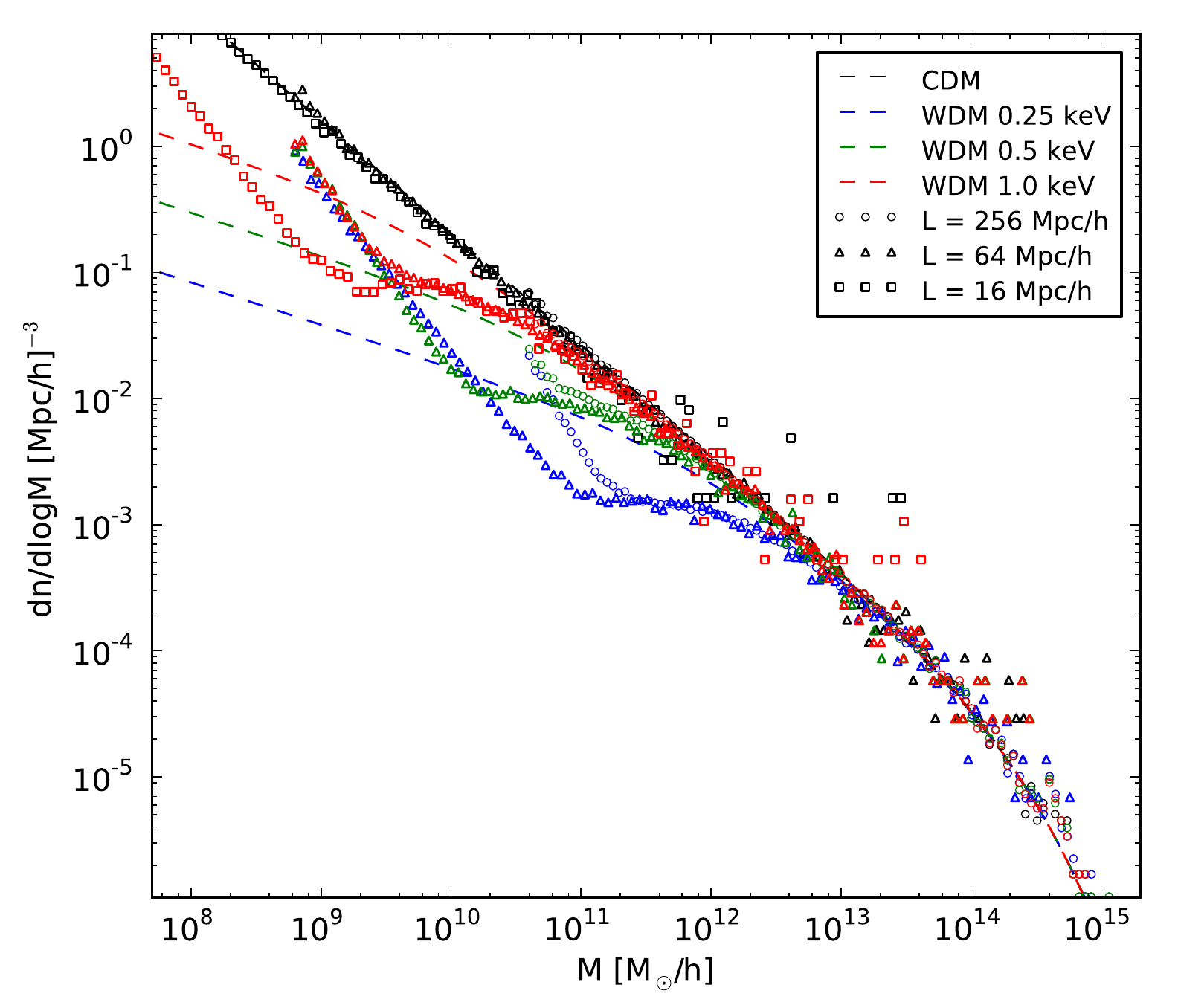}}
\caption{\small{Redshift zero mass functions measured in our
    simulations. Black is the CDM cosmology, blue, green and red are
    WDM cosmologies with particle masses of
    $\mwdm\in\{0.25,0.5,1.0\}\,\keV$. Circles correspond to
    simulations with $L=256\,\Mpc$, triangles to simulations with
    $L=64\,\Mpc$ and squares to simulations with $L=16\,\Mpc$ (each
    with $1024^3$ particles). The low mass upturn in the WDM runs is
    due to spurious structures. The dashed lines represent the usual
    Sheth-Torman mass function.}}\label{simmassfct}
\end{figure}

In the halo mass function the artificial clumps appear as a steep
power-law, which leads to a prominent upturn below a characteristic
mass scale. In \Fig{simmassfct} we plot the mass function of all
simulated boxes together with the usual Sheth-Torman prediction
(dashed lines). The different colours correspond to different
cosmologies (black: CDM, red: $\mwdm=1.0\,\keV$, green:
$\mwdm=0.5\,\keV$, blue: $\mwdm=0.25\,\keV$), the symbols denote the
box size of the simulations (circles: $L=256 \Mpc$, triangles:
$L=64\,\Mpc$; squares: $L=16\,\Mpc$).


The figure clearly shows the dependence of the artifical upturn of the
mass function on the resolution of the simulation. Furthermore, we
note that the mass function of the artificial clumps display a
power-law behaviour. Interestingly, the power-law index of the
artificial mass function does not appear to depend strongly on the
mass of the WDM particle. Instead, it becomes more negative with
decreasing resolution. 

The presence of the artificial clumping is a very serious problem for
numerical simulations, since increasing the resolution becomes
tremendously expensive. \citet{Wang2007} noticed that the effective
converging resolution only goes as $N^{1/3}$ with increasing particle
number $N$. However, as \Fig{simmassfct} shows things are even worse
than that, because the slope of artificial upturn becomes shallower as
the simulation resolution increases, which makes it more difficult to
distinguish between the artificial and real part of the mass
function. Also, it is not possible to simply shrink the box size of
the simulation to increase the mass resolution. Owing to the flatness
of the WDM mass function, small boxes simply do not have enough haloes
to get a statistically meaningful result. The combination of all these
effects makes it incredibly challenging to probe mass scales
significantly below the half-mode mass scale. It is therefore crucial
to attempt to obtain an improved theoretical understanding of how the
mass function behaves at these scales.

Additionally to the resolution problems mentioned above, the exact
scale of the artificial upturn is not always easy to determine. It is
possible that there are still physical haloes in the artificially
driven power-law part of the mass function. On the other hand the
power-law of artificial clumps could extend to higher masses than what
is apparent and change the shape of the mass function at scales well
above the visible upturn.


\subsection{Artificial haloes and environment}\label{correction}

In order to examine the distribution of artificial haloes, we measure
the mass function in different environments. If the visual impression
that the artefacts lie predominantly in filaments is correct, then it
should be possible to extract the halo mass function down to smaller
mass scales by exclusively looking in very underdense void-like environments.

We measure what we will refer to as an ``approximate conditional mass
function'' by imposing a halo isolation criteria.  In practice, we choose
a specific nearest neighbour isolation criterion and only retaining
`isolated' haloes whose nearest neighbour is further away than the 
distance $d$, defined in units of the box size $L$.
This gives an approximate measure of the underdense conditional mass function.
`Non-isolated' haloes on the other hand -- the ones with at least one
neighbour closer than $d$ --
can be used to approximate an overdense conditional mass function.

\begin{figure}
\centering{ \includegraphics[width=8.7cm]{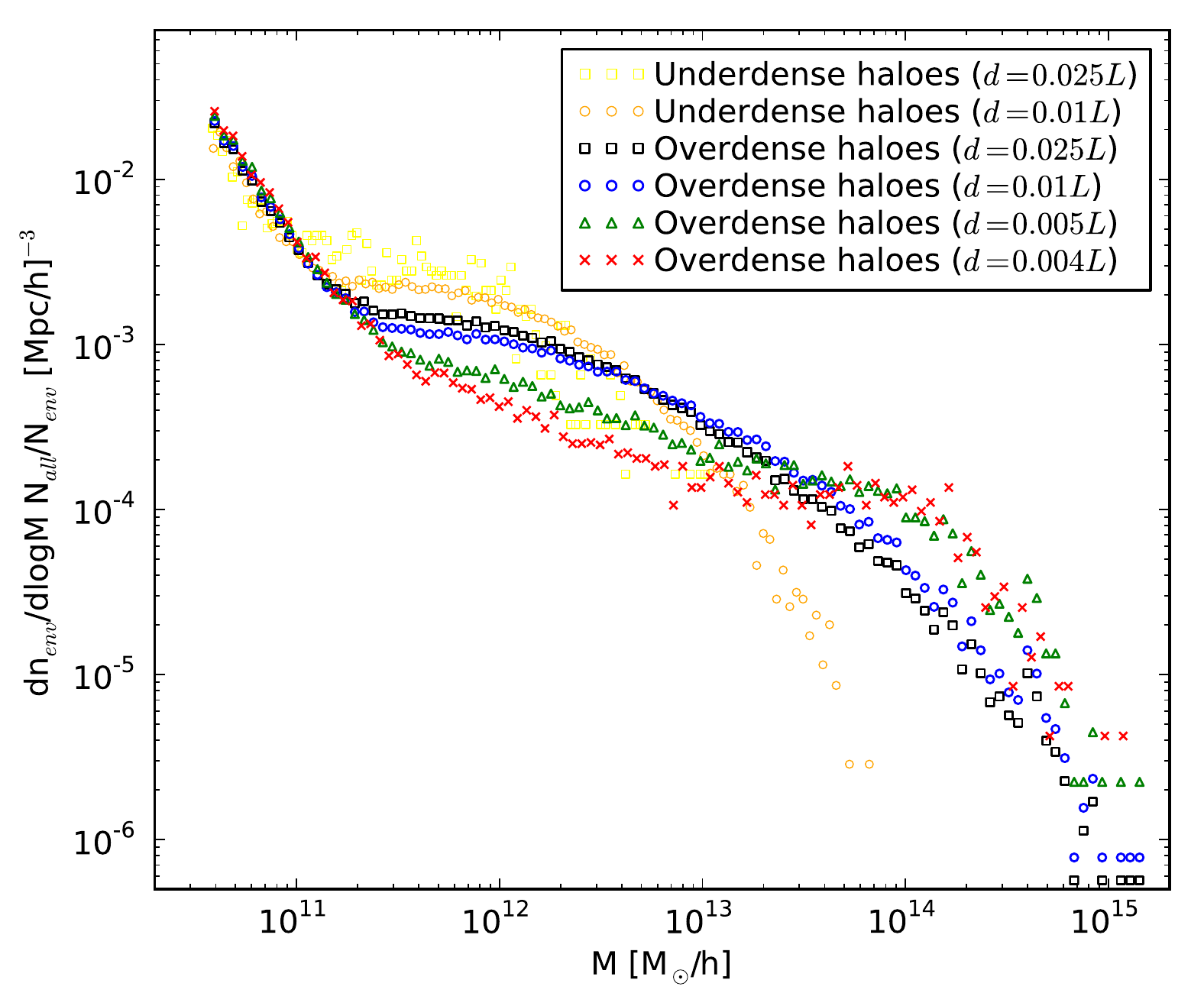}}
\caption{\small{Approximate conditional mass function of underdense
    and overdense environments for the $\mwdm=0.25\,\keV$ model. Haloes in underdense  (overdense) regions have no (at least one) neighbour within the distance $d$, defined in units of the box size $L$. The
    artificial upturn is shifted to smaller (larger) masses in
    underdense (overdense) regions. The power-law of the artificial
    haloes seems to be independent of the
    environment. The halo environment is defined via the distance to the nearest neigbour $d$}.}\label{condmassfct}
\end{figure}

In \Fig{condmassfct} we plot the approximate conditional mass function
of various overdense and underdense environments with varying distance
$d$. The mass function measurements are normalized in a
way that an integration over all mass bins leads to the same effective
halo number -- i.e. the measured mass function in a certain environment $dn_{\rm env}/d\log M$ is multiplied with the ratio $N_{\rm env}/N_{\rm tot}$, where $N_{\rm env}$ and $N_{\rm tot}$ are the total number of haloes in the specific environment and in the whole box, respectively. As expected from visual appearances, the underdense
regions are less contaminated with spurious haloes than the
unconditional mass function, as indicated by the mass scale of the
artificial upturn. Overdense regions on the other hand, have more
artefacts and the artificial upturn therefore happens at larger
masses. The power-law slope of associated with the artificial haloes,
seems to be independent of environment (c.f.~\Fig{simmassfct}, and
even the WDM model).


The fact that the power-law of artefacts is `exposed' up to larger
masses in overdense regions means that it must extend far beyond the
visible upturn even in the unconditional mass function. By extension, it 
also implies that some of the haloes in the artificial upturn must be real. 
In order to get a meaningful approximation of the physical mass function, 
the artificial power-law should therefore be subtracted from the
measurements.  We do this by individually fitting a separate power-law
function to the artificial part of every simulation and then subtract
this power-law function from the measurements. The artificial part is determined via a visible inspection and consists of the data points that are clearly above the upturn, where the shape of the mass function is a pure power law. This approach of subtracting artefacts allows us
to get a meaningful mass function down to the mass scale where the
artificial upturn is a pure power-law. However, it is important to
note that this involves an extrapolation of the spurious halo mass
function to large mass scales.

\begin{figure*}
\centering{ \includegraphics[width=18cm]{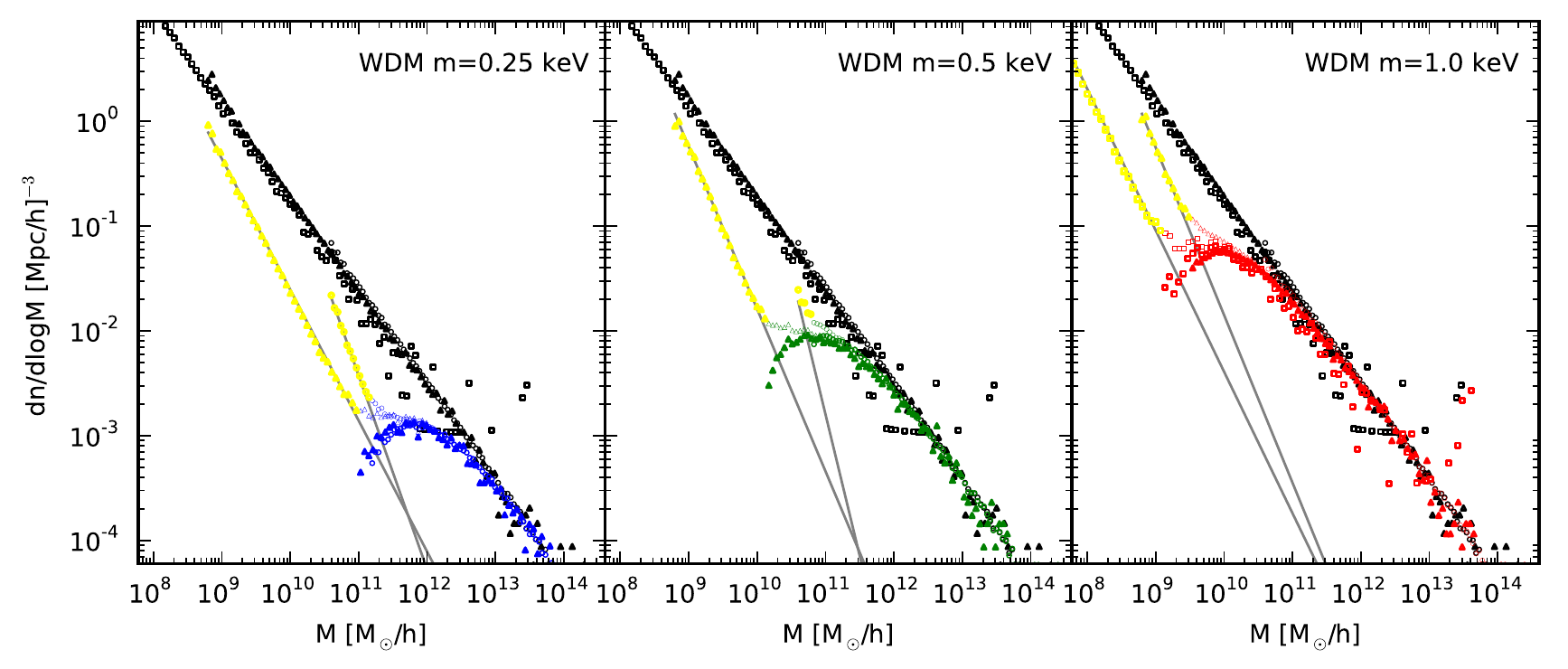}}
\caption{\small{Correction of WDM mass functions for the effects of
    spurious structure formation in the simulations. Left panel: WDM
    with $\mwdm=0.25\,\keV$, middle panel: WDM with $\mwdm=0.5\,\keV$,
    right panel: WDM with $\mwdm=1.0\,\keV$. The CDM measurements have
    been added to every panel for comparison. The faint symbols
    correspond to the original mass function, the bold symbols
    correspond to the corrected mass function. The grey lines are the
    power-laws which are subtracted. The fitting is done over the
    yellow symbols. The WDM mass function decreases towards low masses
    only when this correction is applied.}}\label{mfcorr}
\end{figure*}

In \Fig{mfcorr} the corrected mass function with power-law subtraction
is plotted in solid bold symbols while the faint symbols corresponds
to the original non-corrected mass function. The left panel shows the
measured mass function of the WDM run with $\mwdm=0.25\,\keV$ (blue),
the middle panel the WDM run with $\mwdm=0.5\,\keV$ (green) and
the right panel the WDM run with $\mwdm=1.0\,\keV$ (red). The power
laws, which are subtracted from the original mass function, are
plotted as grey lines and the fitting of these lines is done over the
yellow symbols.


In the following we will only consider the corrected measurement of
the mass function and use it to calibrate our analytical approaches.



\section{Mass Function in a WDM Universe}
The EPS framework \citep{Press1974,Bond1991,Lacey1993,Musso2012} captures many
important features of the end states of structure formation in the CDM
model. In particular, the halo mass function can be defined as
\be\label{massfct}
\frac{dn}{d\log M}=-\frac{1}{2}\frac{\rhob}{M}
f(\nu)\frac{d\log\sigma^2}{d\log M} \ .
\ee
where $f(\nu)$ is the first crossing distribution, $\sigma^2(M)$ the
variance at the mass scale $M$, and $\rhob$ is the average density
of the universe. On assuming uncorrelated random-walks and a collapse
barrier set by the spherical collapse model, the excursion set model
predicts \citep{Press1974,Bond1991}:
\be
f(\nu) = \sqrt{\frac{2\nu}{\pi}}e^{-\nu/2},\hspace{0.5cm}
\nu\equiv\frac{\delta_c^2(t)}{\sigma^2(M)}\ , \label{fnuPS}
\ee
where $\delta_c(t)=1.686/D(t)$ is the linearly extrapolated density
for collapse in the spherical model and $D(t)$ is the growth factor
normalized to be unity at $z=0$. An ellipsoidal collapse barrier gives
\be
f(\nu) = A\sqrt{\frac{2q\nu}{\pi}}
\left[1+(q\nu)^{-p}\right]e^{-q\nu/2}, \label{fnuST}
\ee 
where ${p=0.3}$ and ${A=0.3222}$ \citep{ShethTormen1999}. The third
parameter $q$ is predicted to be one in the ellipsoidal approach, but
Sheth \& Tormen realized that the cluster abundance in simulations is
better matched with the empirical value $q=0.707$.

In this framework all of the sensitivity of the mass function to
cosmology is encoded in the variance of the density perturbations on a
given  scale $R$. The variance can be expressed as
\be\label{var}
\sigma^2(R) = \int \frac{\dk}{(2\pi)^3} P_{\rm Lin}(k) W^2(kR)\ ,
\ee
where $P_{\rm Lin}$ is the linear theory matter power spectrum at 
$z=0$ and $W$ is the Fourier transform of the filter function. Note that in the EPS framework the
predictions are unchanged if we consider, rather than the density
field growing with time and points collapsing when they cross a given
density threshold, the collapse barrier evolves with time and the
field remains static. We adopt this latter convention. Consequently,
it means that the cosmological information encoded in the growth of
the power spectrum is transferred to $\delta_{\rm c}(t)$. Note also
that the value of $\delta_{\rm c}$ itself has a weak cosmology
dependence \citep{Lahav1991,Eke1996}, however in what follows
this shall be neglected.

In the case of perfectly cold dark matter, $\sigma^2(R)$ rises
monotonically towards smaller $R$ and becomes infinite as
$R\rightarrow 0$. On the other hand if free streaming takes place, the
variance becomes constant for small but finite $R$, since the cutoff
in the power spectrum truncates \Eqn{var}.

In order to evaluate \Eqn{massfct} we also require the logarithmic
derivative of the variance with respect to the mass scale, given by\footnote{Here we assume a mass dependence of the form $M\propto R^3$.}
\be\label{dvar}
\frac{d\log\sigma^2}{d\log M}=\frac{2}{3\sigma^2}
\int \frac{\dk}{(2\pi)^3} P_{\rm Lin}(k)W(kR)\frac{dW(kR)}{d\log(kR)}.
\ee
For scales well above the half-mode scale, the behaviour
of this expression is almost independent of the exact choice of the
window function. However, this changes quite dramatically as one
approaches the half-mode scale, whereupon the choice of the window
function dictates the asymptotical behaviour of \Eqn{dvar}, and
thereby determines the shape of the mass function at low masses.


\subsection{Choosing the right window function}\label{windowfunction}
In the presence of particle free streaming the mass function strongly depends on the choice of the window function. Some
common filter functions are the following:
\begin{itemize}
\item {\bf Tophat window:}
This filter has the form of a sphere with radius $R_{\rm TH}$ and 
sharp boundaries in real space and therefore has an obvious mass 
assignment $M_{\rm TH}=4\pi\rhob R_{\rm TH}^3/3$. In Fourier space this 
window has the form:
\be
W_{\rm TH}(y) = \frac{3}{y}\left[\sin y-y\cos y\right],\hspace{1cm} y=kR_{\rm TH}.
\ee
The downside of the spherical tophat filter is that in Fourier space the 
random walk is correlated and the first crossing distribution cannot be 
recovered analytically \citep{Bond1991,Maggiore2010}.
\item {\bf Gaussian window:}
This filter has no sharp boundaries but the characteristic form of a Gaussian (with variance $R_{\rm GA}^2$) in real space. In Fourier space this property is maintained, leading to
\be
W_{\rm GA}(y) = {\rm{e}}^{-y^2/2},\hspace{1cm} y=k R_{\rm GA}.
\ee
A Gaussian window gives somewhat smoother results than a simple tophat window, but it has the drawback of not having a well defined mass. The most common practice to assign a mass is to normalize the filter to one in real space and to integrate over the filter volume. Multiplying the volume with the average density then leads to a filter mass of $M_{\rm GA}=(2\pi)^{3/2}\pi\rhob R_{\rm GA}^3$. The normalization is however arbitrary and this introduces an ambiguity into the mass assignment \citep{Maggiore2010}.
\item {\bf Sharp-$k$ window:}
This filter is defined as a tophat sphere in Fourier space:
\be\label{SKwindow}
W_{\rm SK}(y)=\Theta(1-y),\hspace{1cm}y=kR_{\rm SK},
\ee
and has the very appealing property that the steps of the
random walk are uncorrelated for a Gaussian field.
The drawback is its wiggly shape in real space, leading to contributions on all scales and making it difficult to find a reasonable mass assignment. The same procedure as for the Gaussian filter -- i.e. normalizing the filter and integrating over the enclosed volume -- leads to a divergent integral \citep{Maggiore2010}. Apart from the $M_{\rm SK}\propto R_{\rm SK}^3$ proportionality, the mass assignment is therefore basically unconstrained and needs to be chosen by comparing to simulations (a more detailed discussion on the sharp-$k$ mass assignment is given in \S\ref{massassignment}).
\end{itemize}

We now look at the effect of the different windows on the behaviour of the mass function in a universe including free streaming. The filter specific mass function can be written as
\be\label{massfctW} 
\frac{dn_{\alpha}}{d\log M_{\alpha}}
=\frac{1}{2}\frac{\rhob}{M_{\alpha}}
f_{\alpha}(\nu_{\alpha})\left|\frac{d\log \sigma^2_{\alpha}}{d\log M_{\alpha}}\right|, \hspace{0.5cm}\nu_{\alpha}=\frac{\delta^2_{\rm c}}{\sigma^2_{\alpha}},
\ee
where the index $\alpha$ refers to a particular window function, e.g.
\mbox{$\alpha\in\{\rm TH,\, GA,\, SK\}$}, and so for instance $f_{\rm
  TH}(\nu_{\rm TH})$ is the first crossing distribution computed using
the spherical top-hat window function. 

In the small scale limit below the half-mode scale the variance given in Eq.~(\ref{var}) becomes constant and the asymptotical behaviour of the mass function of Eq.~(\ref{massfctW}) is therefore given by
\be
\lim\limits_{R_{\alpha} \rightarrow 0}{\frac{dn_{\alpha}}{d\log M_{\alpha}}}\propto\frac{1}{R_{\alpha}^{3}}\lim\limits_{R_{\alpha} \rightarrow 0}{\frac{d\log\sigma_{\alpha}^2}{d\log M_{\alpha}}}.
\ee
It is now straight forward to calculate the asymptotical behaviour of the mass function for different filter choices. In the case of a tophat window and equivalently of a Gaussian window we obtain
\be
\lim\limits_{R_{\alpha}\rightarrow 0}{\frac{dn_{\alpha}}{d\log M_{\alpha}}}\propto\lim\limits_{R_{\alpha}\rightarrow 0} R_{\alpha}^{-1}=\infty,\hspace{0.5cm}\alpha=\rm{TH},\rm{GA}
\ee
which means that the mass functions of these filters diverge for small scales. This result is
in contradiction to our understanding of structure formation around
and below the half-mode mass scale. However, In the case of a
sharp-$k$ window the asymptotical behaviour of the mass function is
given by
\ba
\lim\limits_{R_{\rm SK} \rightarrow 0}{\frac{dn_{\rm SK}}{d\log M_{\rm SK}}}&\propto& \lim\limits_{R_{\rm SK} \rightarrow 0}R_{\rm SK}^{-6}P_{\rm Lin}\left(\frac{1}{R_{\rm SK}}\right)\nonumber\\
&\simeq&\lim\limits_{R_{\rm SK} \rightarrow 0}R^{18-n}=0 \ .
\ea
In obtaining the last line we have used
\be
P_{\rm Lin}(1/R_{\rm SK}) \propto R_{\rm SK}^{-n}T^2_{\rm CDM}(1/R_{\rm SK})
T_{\rm WDM}^2(1/R_{\rm SK})\ .
\ee
The asymptotical behaviour of the sharp-$k$ mass function has the correct physical
characteristics -- it becomes strongly suppressed around the scales where free
streaming is dominant. Owing to this and the fact that the first crossing
distribution was derived for the sharp-$k$ window, we will adopt this window 
function throughout the rest of this paper. However, we are still left 
with the task of how we assign mass to this filter function. We shall discuss
this issue in \S\ref{massassignment}.

\subsection{Mass function with sharp-$k$ window}
It is now straight forward to derive an expression for the halo mass function based on the sharp-$k$ window function. Since the logarithmic derivative of the window is
\be
\frac{d W_{\rm SK}}{d\log y}=-y\delta^{D}(1-y),\hspace{1cm}y=kR_{\rm SK},
\ee
where $\delta^D$ is the Dirac delta function, we can evaluate the integral in Eq.~(\ref{dvar}), obtaining
\be\label{dvarSK0}
\frac{d\log\sigma_{\rm SK}^2}{d\log R_{\rm SK}}=-\frac{1}{2\pi^2\sigma_{\rm SK}^2(R_{\rm SK})}\frac{P_{\rm Lin}(1/R_{\rm SK})}{R_{\rm SK}^3}.
\ee
Here we have used the fact that $\Theta(0)=1/2$. We can now implement Eq.~(\ref{dvarSK0}) into the relation
\be\label{massfctSK0}
\frac{dn_{\rm SK}}{d\log M_{\rm SK}}=-\frac{1}{2}\frac{\rhob}{M_{\rm SK}}
f_{\rm SK}(\nu_{\rm SK})\frac{d\log\sigma_{\rm SK}^2}{d\log R_{\rm SK}}\frac{d\log R_{\rm SK}}{d\log M_{\rm SK}},
\ee
to obtain the mass function based on the sharp-$k$ window function.
Note that provided $M_{\rm SK}\propto R_{\rm SK}^3$ the term $d\log M_{\rm SK}/d\log R_{\rm SK}=3$.

With an appropriate mass assignment, Eq.~(\ref{massfctSK0}) gives a very good match to the measurements of both the CDM and the WDM simulations. Especially the flattening and the turnover in the WDM mass function can be described accurately. A direct comparison to the simulations is done in \S\ref{ComparisonSimulations}.


\subsection{Mass assignment}\label{massassignment}
In the previous sections we mentioned that the sharp-$k$ window
function has no well defined mass assigned to its filter scale. This
intrinsic ambiguity can be exploited by choosing a mass assignment
that yields a good agreement with simulations. Owing to the
geometrical scaling of halo mass with radius (at fixed virial halo
density), it is however a reasonable assumption to maintain the
$M\propto R_{\rm SK}^3$ proportionality and we can therefore write
\be
\label{ma} M_{\rm SK}=\frac{4\pi}{3}\rhob\left[cR_{\rm
SK}\right]^3=\frac{M_{\rm TH}}{c^3},
\ee
where $c=R_{\rm TH}/R_{\rm SK}$ is a free constant. \citet{Lacey1993} proposed the
value $c=(9\pi/2)^{1/3}\simeq 2.42$, which can be obtained by
normalizing the filter to one in real space and integrating over the
volume. One part of this integral is however diverging and Lacey \&
Cole set it to zero without any physical motivation \citep[see][for a
  more detailed discussion]{Maggiore2010}.  In this paper, we choose
$c=2.7$, also without physical justification, in order to get an
optimal match with our simulations\footnote{The exact value of $c$ depends on the halo finding method. A mass function based on a spherical overdensity (SO) finder prefers a slightly smaller value than a mass function based on a FoF finder.}.
This value is not only larger than the one from \citet{Lacey1993} but also slightly larger than the value
$c=2.5$ used by \citet{Benson2012}. The difference between our choice
and the one of \citet{Benson2012} comes from the fact that we compare
to the corrected mass function measurements (as explained in
\S\ref{correction}) while they compared to the direct measurement of
the mass function and ignore all spurious haloes above the visible
upturn.


\begin{figure}
\centering{
  \includegraphics[width=8.6cm]{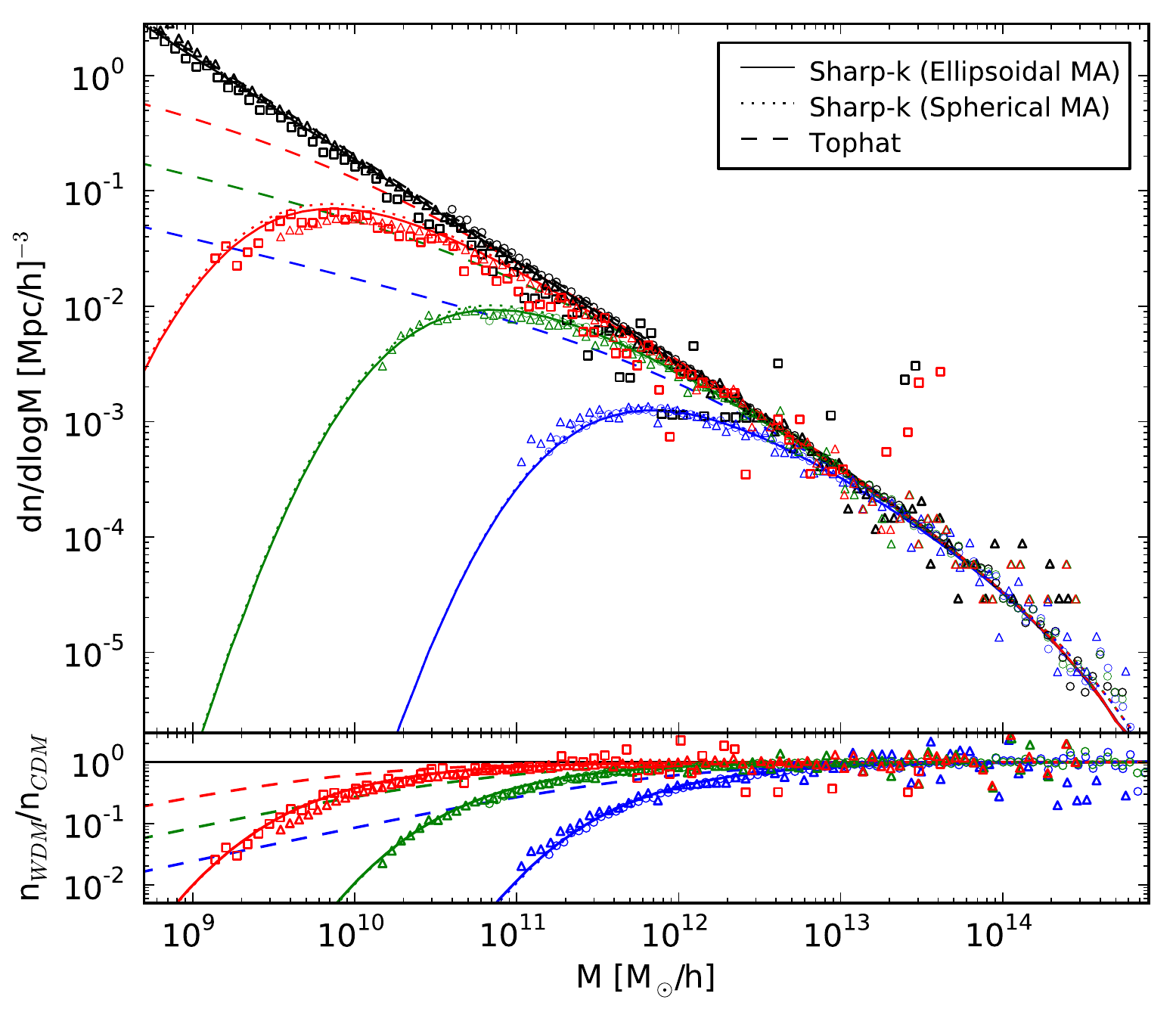}}
\caption{\small{Mass function with tophat filter (dashed) as well as
    sharp-$k$ filter with spherical mass assignment (dotted) and
    ellipsoidal correction (solid). For the spherical mass assignment
    we used $c=2.7$ and $q=1.0$, for the ellipsoidal correction we
    used $c_{\rm e}=2.0$ and $q=0.75$. A tophat window function
    produces a qualitatively incorrect mass function below the free
    streaming scale.}}\label{mfz0}
\end{figure}


Furthermore, in order to obtain a good match with simulations we also
set $q=1$ in \Eqn{fnuST}. This means that with a sharp-$k$ filter we
can use the first crossing distribution, which naturally arises from
the EPS approach with ellipsoidal collapse and we do not need any
further empirical shift of the crossing barrier. Thus, somewhat
suprisingly, we find that neither the rescaling of the first crossing
distribution of \citet{ShethTormen1999} nor the rescaling of the
critical overdensity done by \citet{Benson2012} is necessary. From a
theoretical perspective this means that our sharp-$k$ model is
competitive with the spherical tophat model, since the additional free
parameter from the mass assignment is counterbalanced by the use of a
more natural first crossing distribution.


\subsection{Comparison with simulations}\label{ComparisonSimulations}

The halo mass function predictions from \Eqn{massfctSK0} together with
the mass assignment of \Eqn{ma} can now be compared to the simulated CDM
and WDM scenarios. We do this for the entire mass range spanned by our
simulations \mbox{$M\in\left[10^9,10^{15}\right]\Msol$} and for
redshifts $z=0$ and $z=4.4$. The redshift $z=4.4$ corresponds to the
earliest output for which we have sufficiently good halo number
statistics.

\Fig{mfz0} shows the mass function at $z=0$, where the different dark
matter models are distinguished by colour (black: CDM, blue:
$\mwdm=0.25\, \keV$, green: $\mwdm=0.5\, \keV$, red:
$\mwdm=1.0\,\keV$). The symbols represent the simulation measurements,
with the shape depending on the box size of the simulation (circle:
$L=256\,\Mpc$, triangles: $L=64\,\Mpc$, squares: $L=16\,\Mpc$). The
spherical tophat and sharp-$k$ mass functions are denoted by the
dashed and dotted lines, respectively. The model represented by the
solid line will be discussed in \S\ref{ellipticitycorrection}.


\begin{figure*}
\centering{ 
  \includegraphics[width=5.8cm]{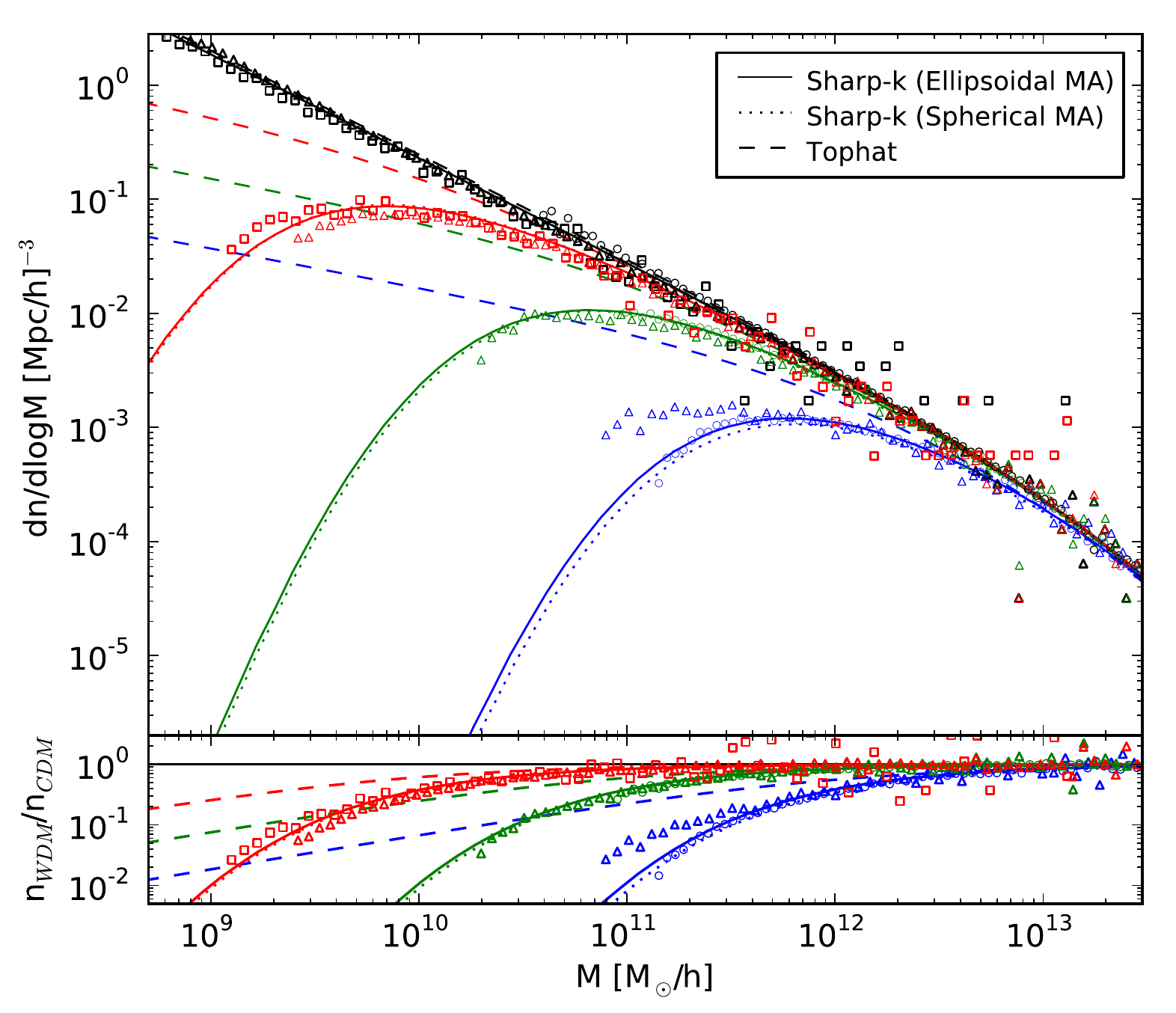}
    \includegraphics[width=5.8cm]{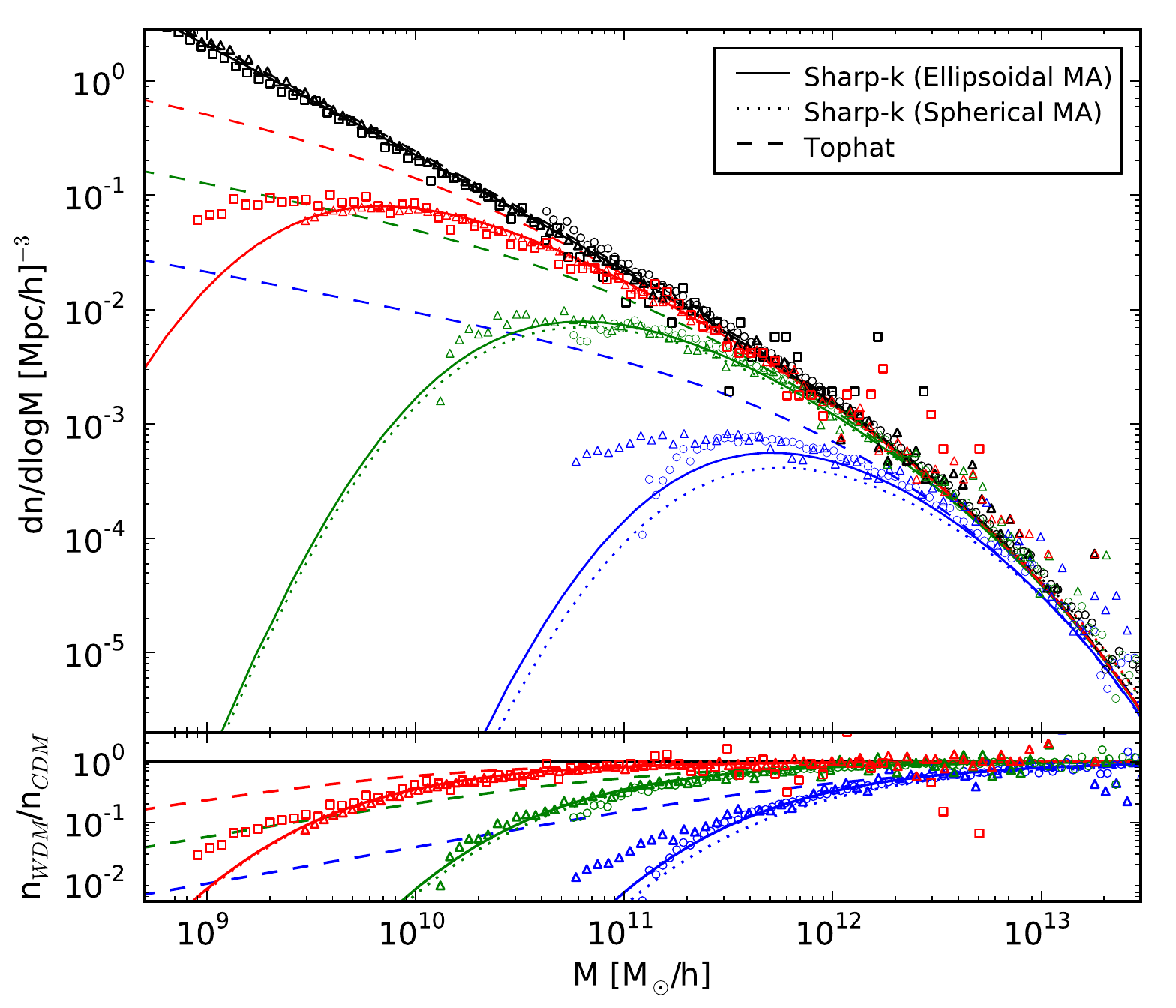}
      \includegraphics[width=5.8cm]{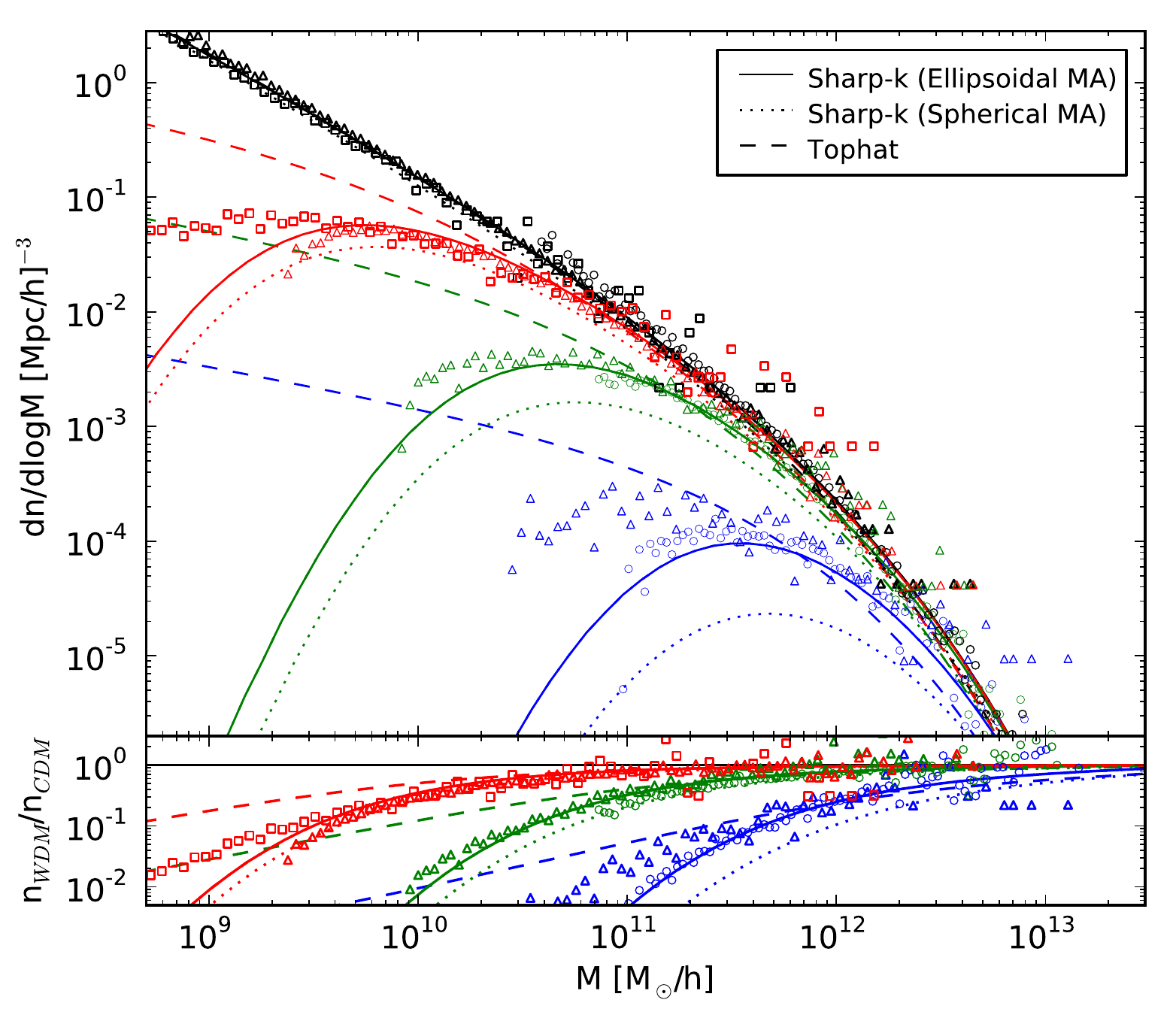}}
\caption{\small{Mass functions at redshift 1.1 (left panel), 2.4
    (middle panel) and 4.4 (right panel). Again, the tophat mass
    function is plotted as dashed line and the sharp-$k$ mass function
    as dotted line while the solid line represents the mass function
    with ellipticity correction, which enables a much better match to
    the simulations at high redshifts. All the labels are the same as
    in Fig.~\ref{mfz0}.}}\label{mfzhigh}
\end{figure*}


For the CDM cosmology, both the spherical tophat and the sharp-$k$
mass functions give very similar predictions. There is a small
difference for very high masses, where the sharp-$k$ mass function
predicts slightly more haloes\footnote{For CDM the difference at large
  masses disappears if a mass assignment of $c=2.42$ \`a la
  \citet{Lacey1993} is used. This is, however, not an option for WDM
  because it leads to an excess in the halo abundance around the half
  mode scale.}. Numerical studies of the high mass end of the mass
function give a halo abundance that lies in between the tophat and the
sharp-$k$ model plotted here
\citep{Reed2007,Bhattacharya2011,Watson2012}. However, and most
importantly, the figure also shows that the predictions for the
sharp-$k$ mass function are in good agreement with the WDM data. On
the other hand, the standard spherical tophat approach significantly
overpredicts the halo abundances at low masses and has the wrong
asymptotical behaviour (see discussion in \S\ref{windowfunction}).

In \Fig{mfzhigh} we plot the evolution of the mass function with
redshift. The left, middle and right panels denote the results at
$z=1.1$, 2.4 and 4.4, respectively. The line and symbol types are as
in \Fig{mfz0}. Unfortunately, as one considers higher redshifts the
sharp-$k$ model predictions begin to systematically underestimate the
measured abundances. This seems to happen as soon as the WDM half-mode mass
scale enters the exponential tail of the mass function. 
At $z=2.4$ and $z=4.4$ the exponential drop-off starts at a mass scale of $10^{11}$ M$_{\odot}$/h and $10^{10}$ M$_{\odot}$/h, respectively. Models with M$^{\rm hm}_{\rm WDM}$ below this drop-off scale reproduce the simulation data well, while the ones above significantly under-predict the amount of haloes.
The effect is most noticeable in the extreme WDM model (blue: $m_{\rm WDM}= 0.25\,
\keV$) at $z=4.4$ (right hand panel), where the discrepancy between
model and data is nearly one order of magnitude. For a WDM model with
more realistic particle mass, $m_{\rm WDM}>1 \keV$, the effect is
small and only becomes significant at very high redshifts.  For
example our most conservative WDM model (red: $m_{\rm WDM}= 1 \,\keV$)
seems to be reasonably well matched at $z=1.1$ and $z=2.4$, and by
$z=4.4$ there is only a slight under-prediction (visible in the
absolute plot but not in the ratio plot).

We conclude that the sharp-$k$ mass function, as given by
\Eqn{massfctSK0} with mass assignment from \Eqn{ma}, seems to provide a
reasonable match to simulation measurements at lower
redshifts. However, the halo abundance of realistic WDM models with
$m_{\rm WDM}>1\, \keV$ is likely to be systematically underestimated
somewhere above $z>5$, when the half-mode mass scale M$^{\rm hm}_{\rm WDM}$ enters the
exponential tail of the mass function.

In the next section we propose that an ellipticity correction of the
sharp-$k$ filter is required. On taking this into account we are able
to obtain predictions that better describe the data for the entire
range of halo masses, redshifts and WDM dark matter particle masses
that we consider in this work.


\subsection{Ellipticity correction}\label{ellipticitycorrection}

One possible explanation for the break down of the sharp-$k$ mass
function predictions for WDM model at high redshifts, could be that
the initial patches of a Gaussian random field are ellipsoidal,
whereas the filter function is spherical and so at best characterises
the effective radius of the patch. The ellipticity of the patches
change both with size and redshift, being more spherical (ellipsoidal)
for large (small) mass scales and at higher (lower) redshift
\citep{Bardeen1986}. In CDM, the effect of the ellipticity can be
folded into the mass assignment. In WDM however, ellipticity becomes
important as soon as the half-mode mass scale (or cutoff scale in
Fourier space) is approached. At this scale only spherical
perturbations survive, while ellipsoidal perturbations do not form
because their shortest axis lies below the half-mode mass scale, where
no power is left.

Correcting for the ellipticity of initial patches only affects the mass 
assignment and is independent of the first crossing distribution. 
The point is to take into account the discrepancy between a spherical 
filter and ellipsoidal patches, which is independent of the effect of 
ellipsoidal collapse, leading to the first crossing distribution of
\citet{ShethTormen1999}.

The distribution of patch shapes for a Gaussian random field was
established by \citet{Bardeen1986}, who found that the expected set of
axis ratios could be described by:
\ba
\frac{a_3}{a_1}(R_{\alpha})& = & 
\sqrt{\frac{1-3e_{\rm m}(R_{\alpha})+p_{\rm m}(R_{\alpha})}
{1+3e_{\rm m}(R_{\alpha})+p_{\rm m}(R_{\alpha})}}\ ,\label{a1a3_1} \\
\frac{a_3}{a_2}(R_{\alpha})& =& 
\sqrt{\frac{1-2p_{\rm m}(R_{\alpha})}{1+3e_{\rm m}(R_{\alpha})+p_{\rm m}(R_{\alpha})}}
\ ,\label{a1a3_2}
\ea
where the ellipsoid axis ratios are defined such that \mbox{$(a_3\le
  a_2\le a_1)$} and where $e_{\rm m}$ and $p_{\rm m}$ are the
ellipticity and prolateness parameters of an average ellipsoidal
density perturbation, both of which depend on the filter scale
$R_{\alpha}$. In Appendix~\ref{AppendixA} we summarize some results
from that work, which are salient for our application. In particular,
the definition of ellipticity and prolateness as well as their
dependence on the filter scale $R_{\alpha}$ and redshift $z$ are
presented.


\begin{figure*}
\centering{
  \includegraphics[width=17.5cm]{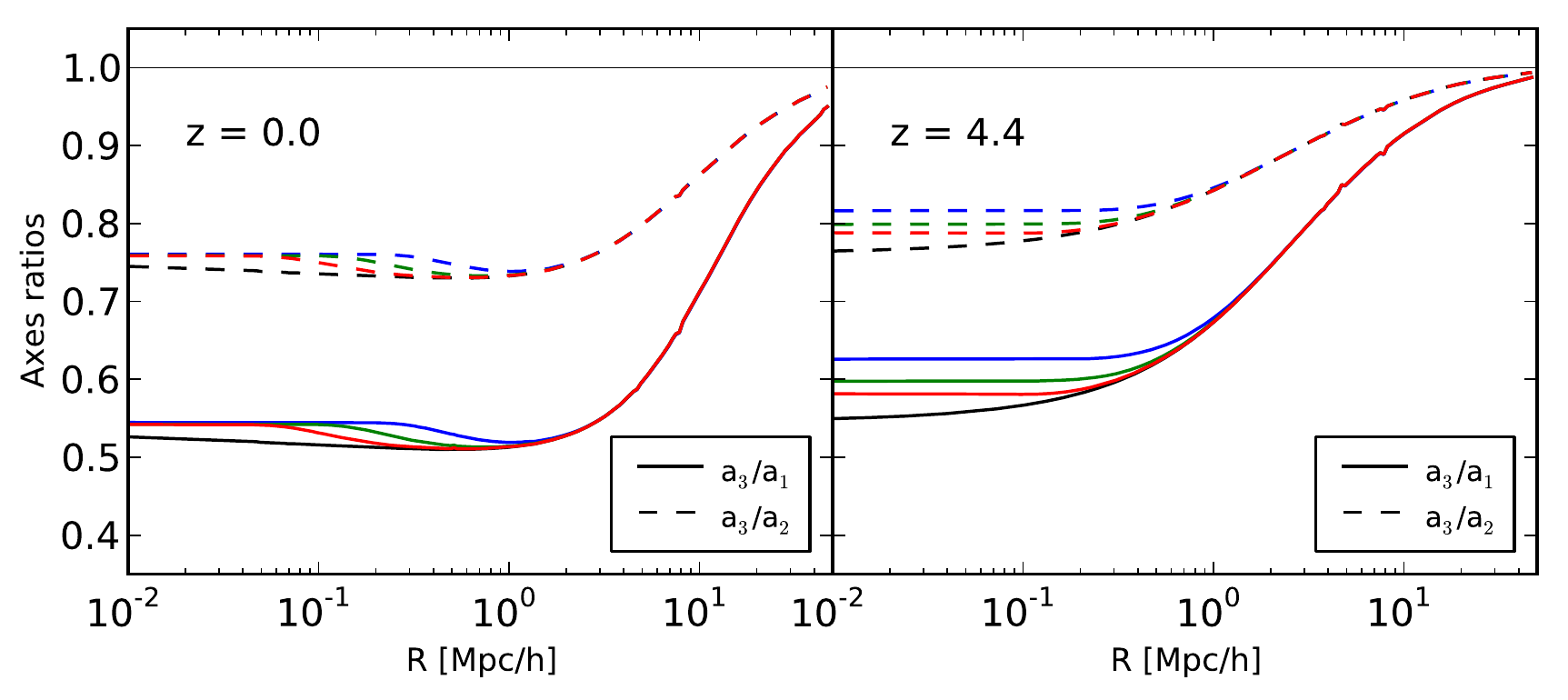}}
\caption{\small{Axes ratios $a_3/a_1$ (solid) and $a_3/a_2$ (dashed)
    at redshift 0 (left panel) and redshift 4.4 (right panel) of the
    initial linear perturbations of the matter density field.} The
    colours represent the different dark matter models (black: CDM;
    blue: $\mwdm=0.25\,\keV$; green: $\mwdm=0.5\,\keV$; red:
    $\mwdm=1.0\,\keV$)}\label{axesratios}
\end{figure*}


In \Fig{axesratios} we show the evolution of the expected axis ratios for a sharp-$k$ filter
as described by Eqns~(\ref{a1a3_1}) and (\ref{a1a3_2}). The left and right panels show the results
at $z=0$ and $z=4.4$, respectively, while the solid and dashed lines
correspond to the ratios $a_3/a_1$ and $a_3/a_2$. The
colours represent the different dark matter scenarios (black: CDM;
blue: $\mwdm=0.25\,\keV$; green: $\mwdm=0.5\,\keV$; red:
$\mwdm=1.0\,\keV$). The patches are clearly more spherical at large
scales and at high redshifts. Also the power spectrum of the WDM model
leads to more spherical patches close to the cutoff scale.

For the case of the sharp-$k$ filter, the connection between the
average radius $R_{\rm SK}$ and the shortest ellipsoidal axis $a_3$
can be obtained by comparing the volume of the filter to the exact
volume of the ellipsoidal patch:
\be\label{Rofa3}
R_{\rm SK}^3=a_1a_2a_3=\left(\frac{a_1}{a_3} \frac{a_2}{a_3}\right)a_3^3=
\left(\xi a_3\right)^3,
\ee
where the ratios $a_1/a_3$ and $a_2/a_3$ and therefore $\xi$ depend on
$R_{\rm SK}$. On combining Eqns~(\ref{a1a3_1}), (\ref{a1a3_2}) and
(\ref{Rofa3}) we get
\ba\label{a3ofR}
&a_3(R_{\rm SK}) &= \frac{R_{\rm SK}}{\xi(R_{\rm SK})},\\
&\xi(R_{\rm SK}) &= \left[\frac{(1+3p_{\rm m}+p_{\rm m})^2}
{(1-3e_{\rm m}+p_{\rm m})(1-2p_{\rm m})}\right]^{1/6}.
\ea
The function $a_3(R_{\rm SK})$ is bijective and can be easily inverted
to obtain $R_{\rm SK}(a_3)$.

Now that we have a relation between the average size of the patch
measured by a spherical filter and the effective smallest ellipsoidal
axis, we can rewrite the mass assignment:
\be\label{ma2}
M_{\rm e}=\frac{4\pi}{3}\rhob\left[c_{\rm e}R_{\rm SK}(a_3)\right]^3,
\ee
where $a_3$ is now used as the reference scale that connects the
linear power spectrum to the final halo abundance. We now conjecture
that as soon as $a_3$ is below the half-mode mass scale, the
corresponding ellipsoid is likely to be erased by the free streaming,
while a spherical filter measuring an average radius $R_{\rm SK}$,
which is still above the half-mode mass scale, will count the
ellipsoid as being existent.

With the replacement of the relevant filter scale from $R_{\rm SK}$ to
$a_3$, the halo mass function can be constructed after the following
recipe:
\begin{enumerate}
\item Compute the first crossing distribution $f_{\rm
  SK}(\nu_{\rm SK})$ as before, using the average patch radius to
  obtain $\nu_{\rm SK}=\delta_c^2/\sigma_{\rm SK}^2(M_{\rm SK})$. No
  adaption to the ellipticity needs to be made here, since shear
  ellipticity is already encapsulated in the EPS approach with
  evolving barrier.
\item Calculate the average ellipticity $e_{\rm m}$ and
  prolateness $p_{\rm m}$ with respect to $R_{\rm SK}$ as described in
  Appendix~\ref{AppendixA}. Determine $\xi (R_{\rm SK})$ and invert
  \Eqn{a3ofR} to obtain $R_{\rm SK}(a_3)$.
\item Determine $M_{\rm e}$ as well as $d\log\sigma_{\rm SK}^2/d\log a_3$ and
  $d\log M_{\rm e}/d\log a_3$.
\item Construct the mass function as follows:
\ba\label{massfctSKell}
\frac{dn_{\rm SK}}{d\log M_{\rm e}}&=&-\frac{1}{2}\frac{\rhob}{M_{\rm e}}f(\nu)
\frac{d\log\sigma_{\rm SK}^2}{d\log a_3}\frac{d\log a_3}{d\log M_{\rm e}},\\
\frac{d\log\sigma_{\rm SK}^2}{d\log a_3}&=&
-\frac{1}{2\pi^2\sigma_{\rm SK}^2(a_3)}\frac{P_{\rm Lin}(1/a_3)}{a_3^3},\label{dvarSK}\\
\frac{d\log M_{\rm e}}{d\log a_3}&=&\frac{3}{\xi(R_{\rm SK})}\frac{dR_{\rm SK}}{da_3}.\label{dlogMSK}
\ea
Note that the last term $d\log M/d\log a_3$ is not exactly 3 anymore
because we have dropped the $M\propto a_3^3$ proportionality in
\Eqn{ma2}.
\end{enumerate}

In order to obtain a good match with the simulations, we take
$c_{\rm e}=2.0$ for the mass assignment as well as $q=0.75$ for the
first crossing distribution presented in \Eqn{fnuST}. This means on
comparison with the usual spherical tophat Sheth-Tormen approach, our
model, with ellipticity correction, has one additional free
parameter. This comes from the undetermined relation between filter
scale and mass inherent to the sharp-$k$ filter (see the discussion in
\S\ref{massassignment}).

The predictions for our sharp-$k$ mass function with an ellipticity
correction are plotted as the solid lines in \Fig{mfz0} and
\Fig{mfzhigh}. At redshift zero the differences between the results
from the sharp-$k$ models with spherical and ellipsoidal mass
assignment are very small. At higher redshifts, however, the corrected
ellipsoidal mass assignment model matches well the estimates from the
simulations. It shows a small under-prediction relative to the highest
resolution simulations, however, the finite volume correction is
comparable to the deviations between model and
data. Furthermore, uncertainties remain in our approximate scheme for
correcting the mass function for spurious haloes.

To summarize we can say that as long as the half-mode mass scale of
the WDM model is situated well above the exponential part of the mass
function, the sharp-$k$ model with spherical mass assignment works
well and no ellipticity correction needs to be done. If the half-mode
mass scale is situated in the exponential part of the mass function,
then the sharp-$k$ model with spherical mass assignment under-predicts
the halo abundance significantly and an ellipticity correction becomes
necessary. The smaller the mass of the WDM particle or the higher the
redshift of interest, the more the half-mode mass scale approaches the
exponential part of the mass function, making an ellipticity correction
essential. For a realistic WDM model with a particle mass around
$\mwdm\sim2\,\keV$, the ellipticity correction is only important for
$z>5$.

The obtained mass function has a self-similar shape for different cutoff scales.
Thus, the half-mode mass scale $M_{\rm WDM}^{\rm hm}$ introduced in \S\ref{sec:powerspectrum} 
defines a relation between the cutoff in $k$-space and the characteristic 
suppression scale in real space and can be used to pin down the peak in the mass function at redshift
zero. Independently of the WDM model, the peak is located at $M_{\rm peak}=0.55 M_{\rm WDM}^{\rm hm}$. Other
interesting scales are also directly related to $M_{\rm WDM}^{\rm hm}$ 
-- for example, the scale where the halo abundance has
dropped by a factor of 10 with respect to the peak --
i.e. $M_{10}\simeq 0.053 M_{\rm WDM}^{\rm hm}$. In principle, with
perfect observations, the rate of decline of the halo mass function
below $M_{\rm peak}$ (or for example the ratio $M_{10}/M_{\rm peak}$) 
might be used to constrain the WDM particle mass.


\section{Predicting the Mass Function of a Neutralino-CDM Universe}

In the standard $\Lambda$CDM picture of the universe, structure
formation spans an enormous range of mass scales from the most massive
galaxy super clusters down to about Earth-mass microhaloes. The reason
for this huge hierarchy is the mass of the favored WIMP particle, the
neutralino, which is of the order of 100 GeV and forms an extremely
cold dark matter fluid at freeze-out with a tiny half-mode mass scale
around one parsec.

In a neutralino-CDM universe the effects of the free streaming are
completely negligible for most astrophysical processes, which serves
also to make their detection a challenge.  For example, microhaloes
are not sufficiently compact to be good targets for microlensing, and
are much too small to gravitationally compress baryons to trigger star
formation (i.e. the cosmic baryon Jeans mass is orders of magnitude
larger). Currently, the most promising means for detecting
microhaloes is via indirect detection through neutralino annihilation
and the resulting by-products. If the microhaloes, or at least their
central cusps, survive the tidal forces of the Milky Way potential,
and if they do not get disrupted by gravitational interactions with
stars, then they are potential gamma-ray sources of the
self-annihilating neutralino \citep{Goerdt2007,Koushiappas2009}.
However, the chances of having a microhalo close enough to the solar
system for easy detection are low, and the overall enhancement of the
signal is moderate \citep{Kamionkowski2008,Schneider2010}.

In the following section, we use pre-existing simulations to test our
model for the neutralino-CDM mass function around the half-mode mass
scale. Finally, we give a prediction for the neutralino-CDM mass
function over the entire range of scales relevant for structure
formation.


\subsection{Comparison with microhalo simulations}
We wish to compare the predictions from our model with results from
$N$-body simulations of the neutralino-CDM model.  There are several
numerical simulation studies presented in the literature
\citep{Diemand2005,Ishiyama2010,Anderhalden2013}, but they tend to
focus on the halo density profile and substructure abundance instead
of the halo mass function. The only measured halo mass function that
we are aware of was reported by \citet{Diemand2005}.

Performing simulations of structure formation in this model is also
incredibly challenging. First, the half-mode mass scale is many
orders of magnitude smaller than for the case of WDM, it being of the
order $M_{\rm CDM}\sim10^{-5}\Msol$. Hence, one requires incredibly
high resolution runs to resolve the relevant mass scale. Furthermore,
owing to the resolution requirement, very small simulation volumes must
be adopted. This means that the initial power spectrum realized in the
simulation volume is close to $P\propto k^{-3}$, and hence nonlinear
structures grow on all scales in the box nearly
simultaneously \citep[e.g.][]{Smith2003,Elahi2009}.

\citet{Diemand2005} tackled the problem by simulating a cosmological
box of size $L = 3\,\kpc$. They then selected an inner region to
re-simulate with a zoom simulation, and this was done for an effective
box size of $L = 60\,h^{-1}{\rm pc}$. The initial power spectrum for
their simulations was based on that from \citet{Green2004}, with a 100
GeV cutoff given by \Eqn{TFcdm} and with $T_{\rm kd}=28$ MeV. The adopted background cosmology
corresponded to WMAP1 with $\Omega_{\rm m}=0.268$, $\Omega_{\rm
  \Lambda}=0.732$, $h=0.71$ and $\sigma_8=0.9$
\citep{Spergel2003}. In order to control the nonlinear evolution of
the box-scale modes, the volume was evolved from an initial start of
$z=350$ and the run halted at $z=26$, by which time a significant
fraction of the mass had collapsed to form microhaloes.

Haloes were selected using a FoF algorithm and were visually
inspected. If a halo was considered to be artificial, it was flagged
and excluded from the mass function estimation.  Note that, because
the accuracy of the visual rejection criteria has not been quantified,
there is a degree of ambiguity left when comparing our predictions
with the simulations.


\begin{figure}
\centering{
  \includegraphics[width=8.5cm]{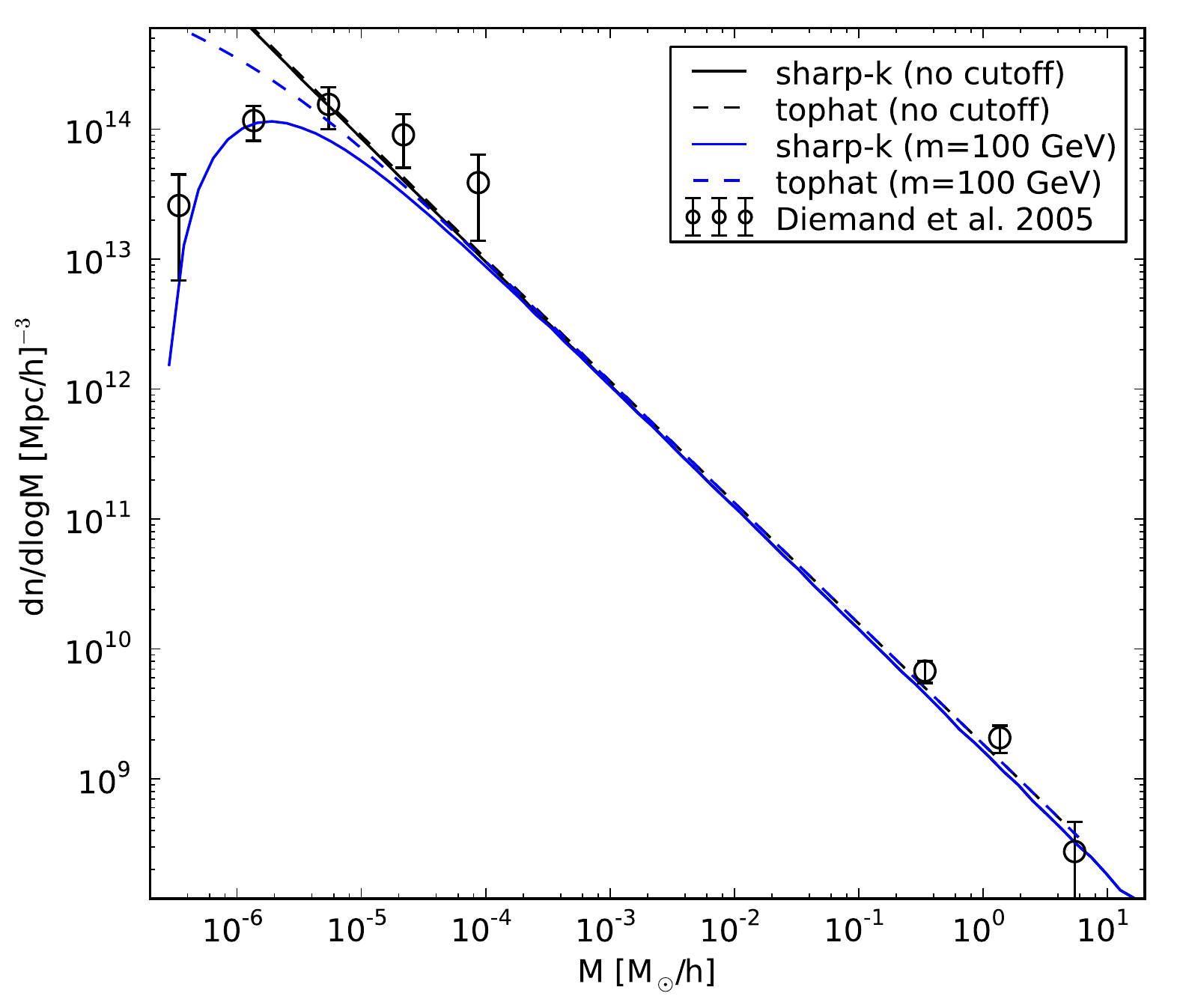}}
\caption{\small{Halo mass function of a 100 GeV neutralino CDM
    scenario with $T_{\rm kd}=28$ MeV and WMAP1 cosmology at redshift
    26. The black dots correspond to the \citet{Diemand2005}
    simulation, where artificial haloes have been removed `by
    eye'. The dashed lines represent the Sheth-Tormen tophat mass
    function, while the solid lines are our sharp-$k$ mass function
    model with ellipticity corrections and the usual parameters
    $c_{\rm e}=2.0$ and $q=0.75$ and yields a small scale decrease
    consistent with the simulation (the simple model using spherical
    mass assignment with $c=2.7$ and $q=1$ gives indistinguishable
    results). Blue lines are with cutoff and black lines
    without.}}\label{cdmcutoff}
\end{figure}


Fig.~\ref{cdmcutoff} presents the measured mass function from the
simulations of \citet{Diemand2005} at $z=26$ as a function of halo
mass. The black circles with error bars denote the estimates from the
simulations: the three data points at the high mass end were derived
from the lower resolution $L=3 \kpc$ simulation, while the five
points at lower masses are measurements from the high resolution zoom
runs. The solid lines show the predictions from our sharp-$k$ space
model mass function with ellipticity correction. The dashed ones show
the results from the spherical tophat Sheth-Tormen model. Black
curves correspond to the CDM model with no cutoff and blue corresponds
to a scenario with 100 GeV neutralino cutoff. Note that the
ellipticity correction has no effect here, since we are far away from
the exponential tail of the mass function, and using the simple
spherical mass assignment with the usual parameters $c=2.7$ and $q=1$
gives indistinguishable results.

The figure clearly shows that our sharp-$k$ mass function model agrees
reasonably well with the simulation data and produces a turnover at
about the right scale. Whilst this is an encouraging result, one
should not over interpret its significance, owing to the poor
statistics and the `by eye' subtraction of artificial haloes.  More
detailed simulations would be necessary to see, for example, if the
slightly steeper cutoff of the neutralino-CDM scenario, versus WDM,
has a visible effect on the shape of the simulated mass function.


\subsection{The complete CDM halo mass function}

We now give a prediction of the redshift zero mass function for a
neutralino-CDM scenario for the WMAP7 cosmological parameters. At
wavenumbers well below the neutralino free-streaming cut-off scale we
adopt the transfer function of \citet{Eisenstein1998}. This is a good
choice for our purposes, since, whilst its accuracy is 2-3\% for large
scale $k$, it asymptotically approaches the exact analytical solution
on small scales (above the cutoff). The neutralino cutoff scale
depends on the mass of the neutralino and the temperature of kinetic
decoupling (c.f.~\Eqns{TFcdm}{kB}). We shall take the mass of the
neutralino to be 100 GeV. The actual decoupling temperature $T_{\rm
  dk}$, however is triggered by collisions with leptons and is model
dependent. \citet{Bertschinger2006} assumed $T_{\rm dk}=22.6$ MeV,
while \citet{Green2005} found $T_{\rm dk}=33$ MeV, based on slightly
different assumptions. In the following we adopt these two values as
benchmarks, but note that other values are possible, depending on the
specific parameters of supersymmetry.

\begin{figure}
\centering{
  \includegraphics[width=8.5cm]{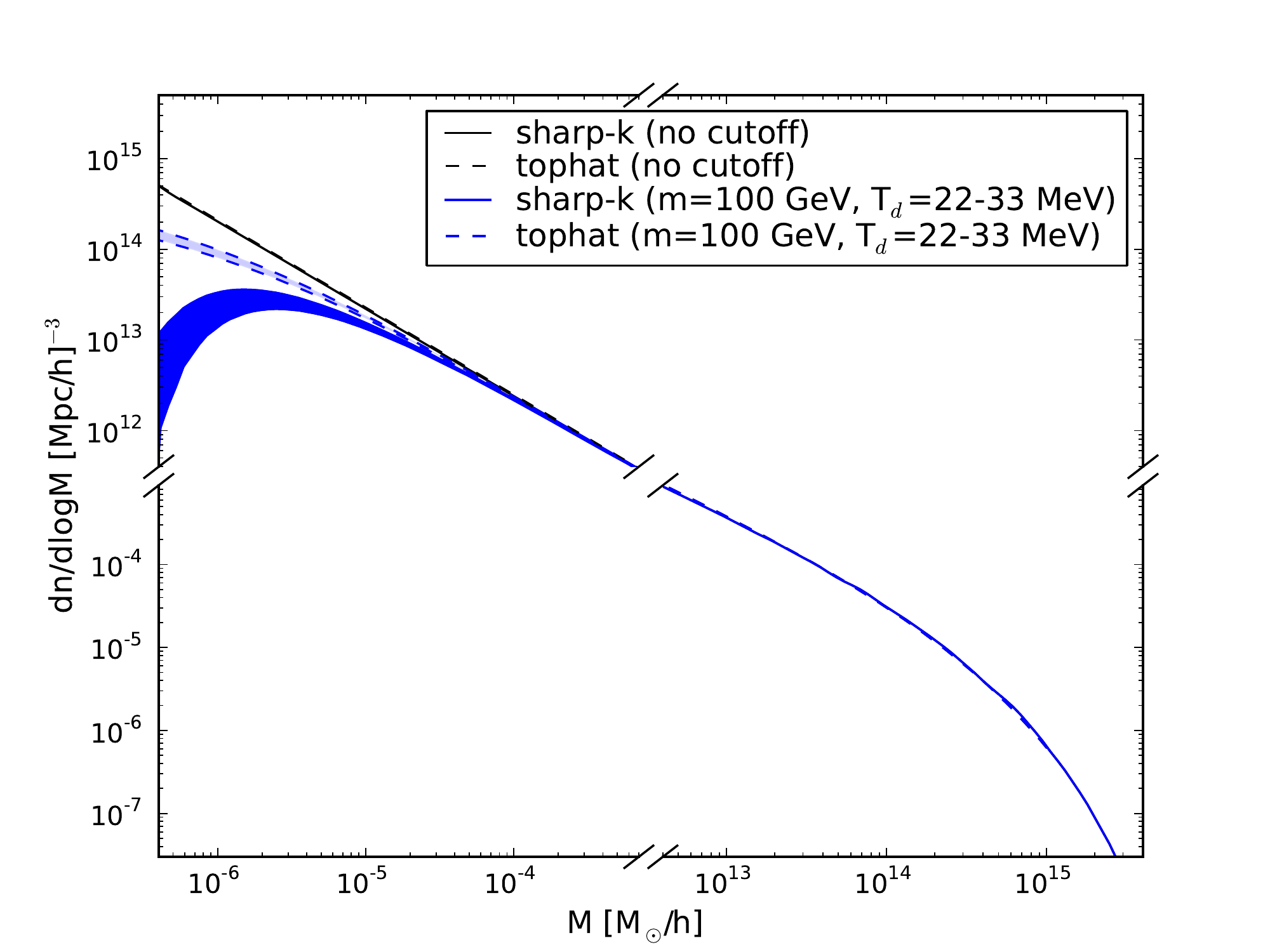}}
\caption{\small{Halo mass function of a 100 GeV neutralino WMAP7
    cosmology at redshift zero. Solid line: sharp-$k$ filter; dashed
    line: Sheth-Tormen tophat filter. The actual half-mode mass scale of a
    100 GeV neutralino depends on the temperature of kinetic
    decoupling which is model dependent. Here we have plotted the two
    benchmark values $T_{\rm dk}=22.6$ MeV and $T_{\rm dk}=33$ which
    delimit the blue shaded regions, where the dark blue region
    represents our new mass function model. The black lines
    corresponds to a model without cutoff.}}\label{mfneutralino}
\end{figure}

\Fig{mfneutralino} shows our prediction for the mass function of dark
matter haloes for a 100 GeV neutralino scenario as a function of halo
mass. Our predictions cover the mass range from the most massive
haloes $M\sim10^{15}M_{\odot}$ until below the scale of one Earth mass
at $M\sim10^{-6}M_{\odot}$. For a better readability of the plot, we
have excised the mass range between $10^{-3}M_{\odot}$ and
$10^{12}M_{\odot}$, where the mass function is essentially a power
law. The blue lines correspond to the model scenario with a 100 GeV
cutoff, while the black lines are without cutoff. The shaded region
enclosed by the dashed lines, denotes the standard spherical tophat
mass function predictions spanned by the two benchmark decoupling
temperatures $T_{\rm dk}=22.6$ MeV and $T_{\rm dk}=33$ MeV. The blue
shaded region enclosed by the solid lines denotes our sharp-$k$ filter
mass function predictions with ellipticity correction and the usual
parameters $c_{\rm e}=2.0$ and $q=0.75$. Again, the simple spherical
mass assignment with $c=2.7$ and $q=1$ gives very similar results.

The prediction of the neutralino-CDM mass function will be useful for estimations of the neutralino annihilation rate in the local universe.


\section{Conclusions}\label{conclusions}
In this paper we have developed a simple analytical model for the halo mass function in the presence of particle free streaming. This is an important task since numerical simulations produce artificial clumping at the scale corresponding to the power spectrum cutoff (half-mode scale) and it is currently impossible to numerically test structure formation below this scale. In a WDM scenario, however, it is crucial to understand the physics of structure formation below the half mode scale, since it is precisely on these scales where the alternative dark matter scenario can be either ruled out or confirmed.

Our method is based on the assumption that there is a direct mapping between the linear power spectrum and the final distribution of haloes, even in the presence of a power spectrum cutoff. In other words, this means that we assume the Extended Press-Schechter (EPS) approach to work for all cosmologies independent of the shape of the power spectrum. We note that this assumption remains unverified; it is for example imaginable that an EPS approach based on linear perturbation theory is not sufficient as soon as the power spectrum deviates from its quasi power-law shape, where EPS is much better verified against cosmological simulations. Instead, an approach with higher order perturbation could be necessary, or the EPS method could break down completely.  On the longer term, it will be therefore indispensable to develop more accurate numerical methods capable of modelling the physics of gravitational structure formation on all scales without producing artificial clumping.

In the following, we list some key findings of this paper and discuss their range of applicability:
\begin{itemize}
\item In order to circumvent the problem of artificial clumping in our simulations we proposed an approximate method to subtract numerical artefacts from the halo mass function. The method is based on the observation that the mass function of artificial haloes has the form of a steep power-law that continues well above the mass range where artefacts dominate structure formation. This power-law is then simply subtracted from the measurements, a correction that makes the mass function turn over around the half-mode scale and decrease towards small masses, just as is expected from theoretical arguments. The power-law subtraction works surprisingly well at low redshift, where the corrected measurements of different box sizes are well-converged. At high redshift, however, the method seems to be less accurate as reflected by a poorer convergence between different box sizes, which appears to be caused in part because the artificial tail of the mass function is no longer such a clear power-law.
\item We have constructed a simple model for the halo mass function based on a sharp-$k$ filter with constant $M\propto R^3$ mass assignment and a first crossing distribution from ellipsoidal collapse. It has the same number of free parameters as the standard Sheth-Tormen approach and gives accurate predictions for CDM as well as WDM scenarios at low redshift. At high redshift, however, as soon as the WDM half mode scale enters the exponential tail of the first crossing distribution, this simple method breaks down leading to an under-prediction of the halo abundance (Fig.~\ref{axesratios}, dotted lines). For a canonical 2 keV WDM candidate this happens at $z>5$.

\item The breakdown of the simple model at high redshift comes from the fact that ellipsoidal patches are smoothed with a spherical window function, something that becomes important in WDM around the half-mode scale, where patches with high ellipticity cannot survive the free streaming. To improve upon this model, we include the effect of the patch ellipticity by taking the shortest ellipsoidal axes as the reference scale instead of the average radius that results when a spherical window function is applied.  With this correction, our model yields an accurate mass function for CDM and WDM at all redshifts tested by our simulations.

%
\item As a further application we used our model to predict the behaviour of the neutralino-CDM mass function, where the half-mode scale lies at an extremely small scale and where the power spectrum cutoff has a somewhat different form. A comparison to simulations from \citet{Diemand2005} shows reasonable agreement, but more precise simulations are necessary to test our mass function model in more detail.
\end{itemize}

The viability of the sharp-$k$ mass function at small scales well below the half mode scale still needs to be tested against simulations. 
This requires however a completely new numerical approach, since all common numerical schemes produce artefacts at the relevant scales, and this consists of a formidable scientific challenge.

\section*{Acknowledgements}
We thank J\"urg Diemand for providing us with the data points of \Fig{cdmcutoff}. AS acknowledges a fellowship from the European CommissionÕs Framework Programme 7, through the Marie Curie Initial Training Network CosmoComp (PITN-GA-2009-238356). RES was supported by the ERC advanced grant 246797 GALFORMOD from the European Research Council. All simulations were performed on the SuperMUC cluster in Munich and the CSCS cluster `Monte Rosa' in Lugano.

\section*{Download Code}
A code which calculates the sharp-$k$ mass function with or without ellipsoidal correction can be downloaded at:
{\tt http://www.phys.susx.ac.uk/{\raise.17ex\hbox{$\scriptstyle\sim$}}as721}.



\input{mnras1.bbl}

\appendix
\section{Geometry of Patches}\label{AppendixA}
Patches around peaks of a Gaussian random field have an ellipsoidal form in the neighborhood around the peak. This observation was first done in the seminal paper of \citet{Bardeen1986}, and it holds independently of the form of the underlying power spectrum. We will now summarize some of the results from \citet{Bardeen1986}, and detail how the ellipsoidal axes are connected to the underlying power spectrum.

Around a local peak the density can be approximated by a Taylor expansion
\be\label{patchdensity}
\delta(r)=\delta(0)-\sum \lambda_i \frac{r_i^2}{2}, \hspace{1cm}\lambda_i=\frac{\left[2\delta(0)-d\right]}{a_i^2},
\ee
where the constant density $d$ defines an ellipsoidal patch and connects the eigenvalues $\lambda_i$ of the tensor $\zeta_{ij} = \partial_i\partial_j\delta$ to the semi major axes of an ellipsoid $a_i$. The geometry of an ellipsoidal patch is characterized by the ellipticity and prolatness parameters
\be\label{ellipticity}
e=\frac{\lambda_3-\lambda_1}{2(\lambda_1+\lambda_2+\lambda_3)},\hspace{1cm}p=\frac{\lambda_3-2\lambda_2 + \lambda_1}{2(\lambda_1+\lambda_2+\lambda_3)}.
\ee
From \Eqns{patchdensity}{ellipticity} it is now straight forward to derive the ratios of the ellipsoidal axes
\be\label{a1a3AP}
\frac{a_3}{a_1}=\sqrt{\frac{1-3e+p}{1+3e+p}},\hspace{1.0cm}\frac{a_3}{a_2}=\sqrt{\frac{1-2p}{1+3e+p}},
\ee
that are used in the main text of this work. The ellipticity corrected mass function given by the Eqns~(\ref{massfctSKell}), (\ref{dvarSK}) and (\ref{dlogMSK}) can therefore be uniquely determined, provided we know the distribution of ellipticity and prolateness in our cosmology. 

\citet{Bardeen1986} found analytical prescriptions of the probability distributions $P(e,p|x)$ as well as $P(x|\nu)$ (Eq. 7.6 and 7.5 in their paper), where $x=(\lambda_1+\lambda_2+\lambda_3)/\sigma_2$ is the sum of the eigenvalues of the tensor $\zeta_{ij}$ divided by the second spectral moment. The spectral moments are defined as
\be\label{varj}
\sigma^2_j(R) = \int \frac{\dk}{(2\pi)^3} k^{2j} P_{\rm Lin}(k) W^2(kR),
\ee
and $\sigma_0$ is simply the variance defined in \Eqn{var}. The two probability distributions can be combined to obtain
\be\label{probdist}
P(e,p|\nu)=\int dx P(e,p|x)P(x|\nu),
\ee
which connects the ellipticity and prolateness parameters to the peak height $\nu$ and therefore the underlying power spectrum.

The distribution $P(x|\nu)$ is sharply peaked and has a maximum at
\be\label{xm}
x_m=\gamma\nu + \frac{3(1-\gamma^2) + (1.1-0.9\gamma^4){\rm e}^{-\gamma(1-\gamma^2)(\gamma\nu/2)^2}}{[3(1-\gamma^2)+0.45+(\gamma\nu/2)^2]^{1/2}+\gamma\nu/2}
\ee
where $\gamma=\sigma_1^2/(\sigma_0\sigma_2)$ \citep[Eq. 6.17]{Bardeen1986}. We ignore the distribution around $x_{\rm m}$ and just set $P(x|\nu)=\delta_{\rm D}(x-x_{\rm m})$, leading to to the simplified distribution $P(e,p|\nu)= P(e,p|x_{\rm m})$.

\citet{Bardeen1986} found that $P(e,p|x_{\rm m})$ is approximately Gaussian (for large x) and has a maximum at
\be\label{exrelation}
e_{\rm m}=\frac{1}{(5x_{\rm m}^2+6)^{1/2}},\hspace{1cm}p_{\rm m}=\frac{30}{(5x_{\rm m}^2+6)^2}.
\ee
It is now straight forward to derive average axes ratios by substituting \Eqn{exrelation} into \Eqn{a1a3AP}.
The result of this calculation is given in \Fig{axesratios} (see main text for more information).

\section{Subtraction of artificial haloes at higher redshifts}
\begin{figure*}
\centering{
  \includegraphics[width=16cm]{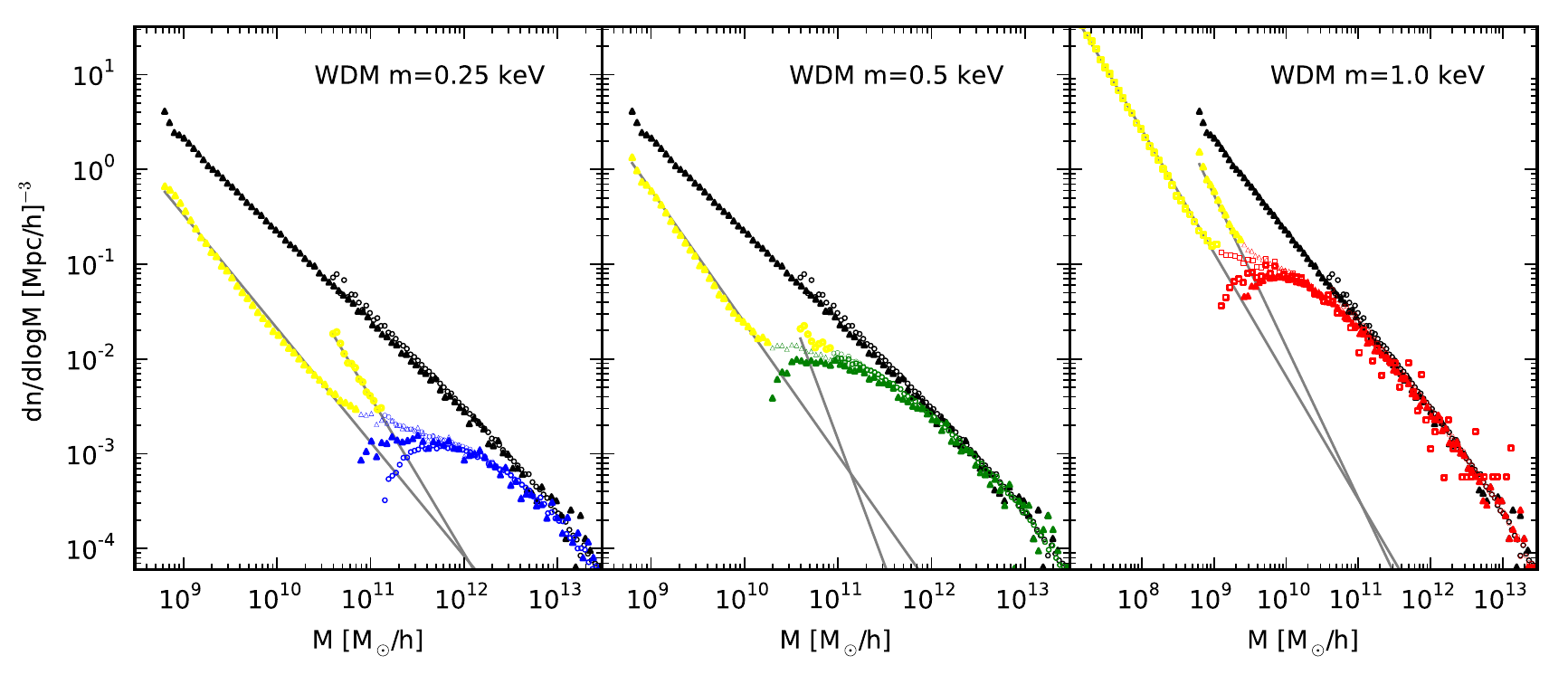}\\
    \includegraphics[width=16cm]{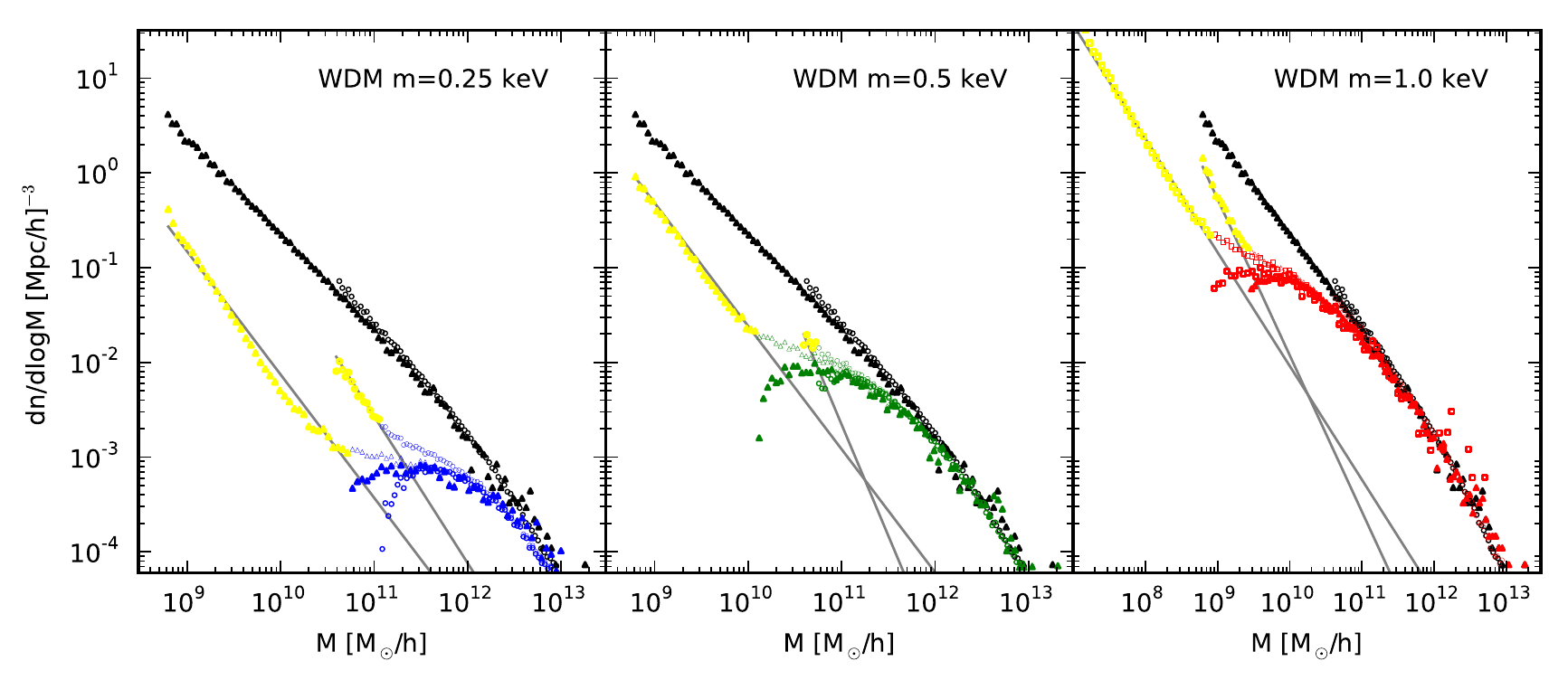}\\
      \includegraphics[width=16cm]{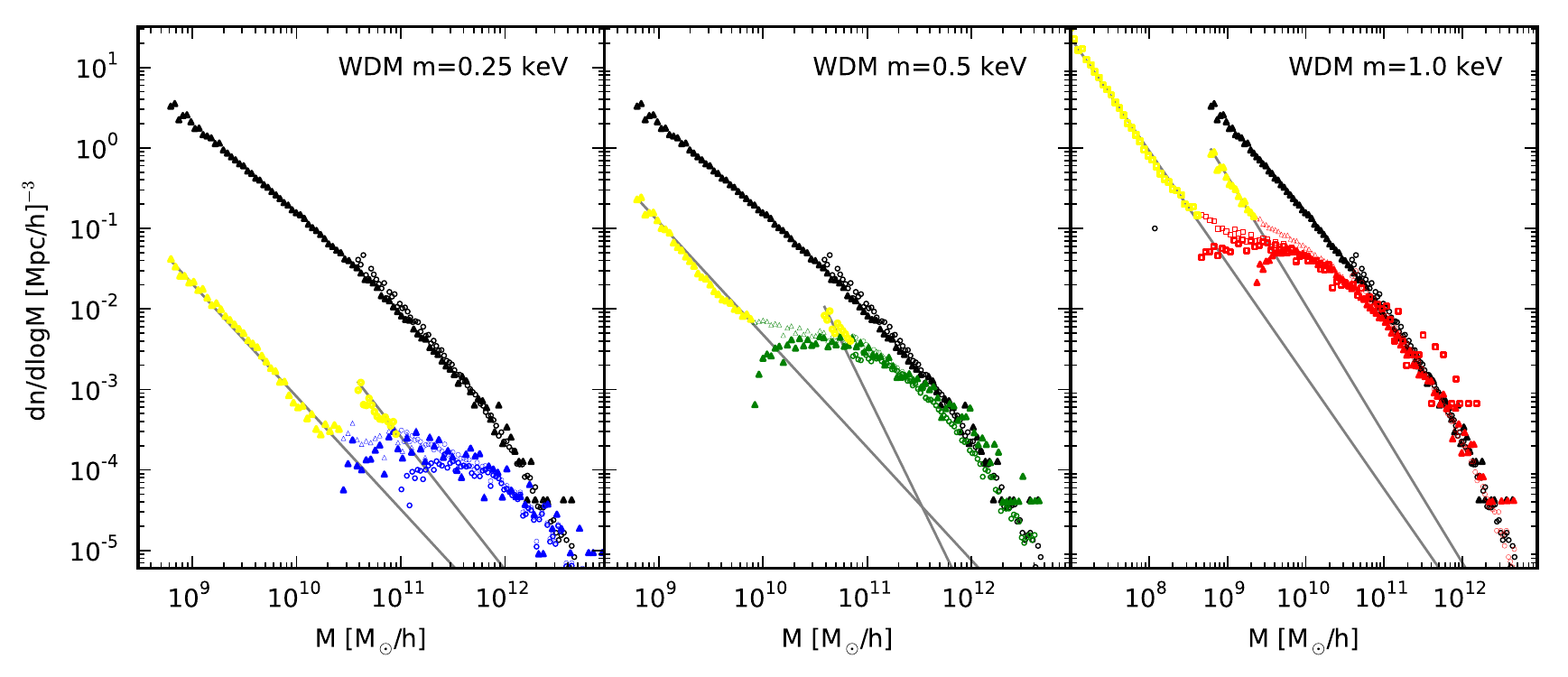}}
\caption{\small{Correction of WDM halo abundance at higher
    redshifts. From top to bottom: $z=1.1$, $z=2.4$ and $z=4.4$. From
    left to right: WDM with $\mwdm=0.25\,\keV$, WDM with
    $\mwdm=0.5\,\keV$ and WDM with $\mwdm=1.0\,\keV$. The CDM
    measurements have been added to every panel for comparison. The
    faint symbols correspond to the original mass function, the bold
    symbols correspond to the corrected mass function. The grey lines
    are the power-laws which are used for the subtraction. The fitting is done over
    the yellow symbols.}}\label{mfcorrzhigh}
\end{figure*}

In the main text we discussed the power-law subtraction of the artificial haloes in detail, and we plotted the utilised power-law fit in \Fig{mfcorr}. Here we will give the same plots for the correction of halo abundance at higher redshift (\Fig{mfcorrzhigh}).

As a matter of fact, the higher the redshift the less the upturn has the shape of a single power-law. This makes it more difficult to do a proper fit and leaves space for ambiguity in the model building. 


\end{document}

%% file: defs.tex
\newcommand\Eqn[1]     {Eq.\,(\ref{#1})}
\newcommand\Eqns[2]    {Eqs\,(\ref{#1}) and~(\ref{#2})}

\newcommand\Fig[1]     {Fig.\,{\ref{#1}}}

\newcommand\nn         {\nonumber}

\def\mnras{{Mon.~ Not.~ R.~ Astron.~ Soc.~}}

\def\prd{{Phys.~ Rev.~ D.~}}

\def\apj{{Astrophys.~ J.~}}
\def\apjs{{Astrophys.~ J.~ Suppl.~}}
\def\apjl{{Astrophys.~ J.~ Lett.~}}

\def\mnras{{MNRAS}}

\def\prd{{PRD}}

\def\apj{{ApJ}}
\def\apjs{{ApJS}}
\def\apjl{{ApJL}}
\def\aap{{A\&A}}

\def\physrep{{Phys.~ Rep.~}}
\def\jcap{Journal of Cosmology and Astro-Particle Physics}

\newcommand{\be}{\begin{equation}}
\newcommand{\ee}{\end{equation}}
\newcommand{\ba}{\begin{eqnarray}}
\newcommand{\ea}{\end{eqnarray}}

\def\pp1{{\prime}}
\def\pp2{{\prime\prime}}

\def\2D{{\rm 2D}}

\def\1Loop{{\rm 1Loop}}

\def\rhob{\bar{\rho}}

\def\Msol{h^{-1}M_{\odot}}
\def\kpc{\, h^{-1}{\rm kpc}}
\def\Mpc{\, h^{-1}{\rm Mpc}}

\def\kMpc{\, h \, {\rm Mpc}^{-1}}

\def\dk{{\rm d}^3{\bf k}}

\def\la{\mathrel{\mathpalette\fun <}}
\def\ga{\mathrel{\mathpalette\fun >}}
\def\fun#1#2{\lower3.6pt\vbox{\baselineskip0pt\lineskip.9pt
        \ialign{$\mathsurround=0pt#1\hfill##\hfil$\crcr#2\crcr\sim\crcr}}}

\def\1H{{\rm 1H}}
\def\2H{{\rm 2H}}

\def\keV{{\rm keV}}
